\documentclass[12pt]{article}
\usepackage[text={18cm,23cm},top=3cm]{geometry}  % This line determines the page layout.
\usepackage{amsmath,amsfonts,latexsym,mathtools,mathrsfs}
\usepackage{graphicx, subfigure}
\usepackage[hidelinks]{hyperref}
\usepackage{xcolor}
\usepackage{titlesec}
\usepackage[titletoc,toc,title]{appendix}

\graphicspath{{Figures/}}

\numberwithin{equation}{section}

\DeclarePairedDelimiter{\abs}{\lvert}{\rvert}

\newcommand{\lb}{\left (}
\newcommand{\rb}{\right )}
\newcommand{\lset}{\left \{}
\newcommand{\rset}{\right \}}
\newcommand{\lsq}{\left [}
\newcommand{\rsq}{\right ]}

\newcommand{\eqtext}[1]{\quad \text{#1} \quad}
\newcommand{\RA}{\quad \Rightarrow \quad}

\newcommand{\sechn}[2]{\mathrm{sech}^{#1} \lb #2 \rb}

\newcommand{\dd}[1]{\; \mathrm{d} #1}
\newcommand{\diff}[2]{\frac{\mathrm{d} #1}{\mathrm{d} #2}}
\newcommand{\diffn}[3]{\dfrac{\mathrm{d}^{#1} #2}{\mathrm{d} #3^{#1}}}

\renewcommand{\O}[1]{O \lb #1 \rb}

\begin{document}

\baselineskip=15pt
\vspace{-2.5cm}
\title {Weakly-nonlinear solution of coupled Boussinesq equations and radiating solitary waves}
\date{}
\maketitle
\vspace{-22mm}
\begin{center}
{\bf K.R. Khusnutdinova$^{*}$
\footnote{Corresponding author: K.Khusnutdinova@lboro.ac.uk}, M.R. Tranter$^{*}$} \\[2ex]
$^{*}$ Department of Mathematical Sciences, Loughborough University, \\
Loughborough LE11 3TU, UK.
\vspace{4mm}

\end{center}

\abstract{Weakly-nonlinear waves in a layered waveguide with an imperfect interface (soft bonding between the layers) can be modelled using coupled Boussinesq equations. We assume that the materials of the layers have close mechanical properties, in which case the system can support radiating solitary waves. We construct a weakly-nonlinear d'Alembert-type solution of this system, considering the problem in the class of periodic functions on an interval of finite length. The solution is constructed using a novel multiple-scales procedure involving fast characteristic variables and two slow time variables. Asymptotic validity of the solution is carefully examined numerically. We also discuss the limiting case of an infinite interval for localised initial conditions.  The solution is applied to study interactions of radiating solitary waves.

\bigskip
\bigskip
\bigskip


\newpage

\section{Introduction}

Low-frequency wave propagation in solids  is relevant to a large number of modern applications (see, for example, \cite{S_book, P_book, PS, IZKS, PTE, ADO, ADKM} and references therein).
Long longitudinal bulk strain solitary waves in elastic waveguides can be modelled using Boussinesq-type equations \cite{S_book, P_book, OS} (see also \cite{ PTE, NS,  DF, EKS, GKS}). The stability of bulk strain solitons makes them an attractive candidate for the introscopy of  layered structures, in particular delamination, in addition to the existing methods \cite{KS, DKSS12, KT15, KT17, BBS}.

The dynamical behaviour of layered structures depends both on the properties of the bulk material in the layers, and on the type of the bonding between the layers. If the materials of the layers have similar properties and the bonding between the layers is sufficiently soft (``imperfect interface"), then the bulk strain soliton is replaced with a {\it radiating solitary wave}, a solitary wave with a co-propagating oscillatory tail \cite{DKSS12, KSZ, GKM}. 

Long nonlinear longitudinal bulk strain waves in a bi-layer with a sufficiently soft bonding can be modelled with a system of coupled regularised Boussinesq (cRB) equations \cite{KSZ} (given below in non-dimensional and scaled form):
\begin{align}
&u_{tt} - u_{xx} = \epsilon \lsq \frac{1}{2} \lb u^2 \rb_{xx} + u_{ttxx} - \delta \lb u - w \rb \rsq, \label{ueq} \\
&w_{tt} - c^2 w_{xx} = \epsilon \lsq \frac{\alpha}{2} \lb w^2 \rb_{xx} + \beta w_{ttxx} + \gamma \lb u - w \rb \rsq, \label{weq}
\end{align}
where $\alpha$, $\beta$, $\delta$, $\gamma$ are coefficients depending on the mechanical and geometrical properties of a waveguide, $\epsilon$ is a small amplitude parameter, and $c$ is the ratio of the characteristic linear wave speeds in the layers. We assume that  the materials of the layers have close mechanical properties, and therefore $c^2 - 1 = \mathcal{O}(\varepsilon)$.

We consider the initial-value (Cauchy) problem, and the initial conditions are written as
\begin{align}
&u(x,0) = F_{1}(x), \quad u_{t} (x,0) = V_{1}(x), \label{uIC} \\
&w(x,0) = F_{2}(x), \quad  w_{t} (x,0) = V_{2}(x). \label{wIC}
\end{align}
%Let us first study the dispersion relation of the system \eqref{ueq} - \eqref{weq}. 

%The linear waves 
%proportional to $e^{i(kx-\omega t)}$ 
The dispersion relation is given by the bi-quadratic equation \cite{KSZ}:
\begin{align}
&\omega^4 \lb 1 + \epsilon \beta k^2 \rb \lb 1 + \epsilon k^2 \rb \notag \\
&-~ \omega^2 \lsq \lb 1 + \epsilon \beta k^2 \rb \lb \epsilon \delta + k^2 \rb + \lb 1 + \epsilon k^2 \rb \lb \gamma \epsilon + c^2 k^2 \rb \rsq + \epsilon \lb \gamma + \delta c^2 \rb k^2 + c^2 k^4 = 0,
\label{DispRel}
\end{align}
where $k$ is the wave number, and $\omega$ is the wave frequency.
The two roots of this equation give two modes: acoustic and optical. The acoustic mode has the dispersion relation $\omega = \omega_a(k)$ satisfying the two asymptotic approximations
\begin{equation}
\omega_a^2(k) = \frac{\gamma + \delta c^2}{\gamma + \delta} k^2 + \mathcal{O}(k^4) \quad \mbox{\rm as} \quad k \to 0 \eqtext{and} \omega_a^2(k) = \frac{c^2}{\epsilon \beta} + \mathcal{O}(k^{-2}) \quad \mbox{\rm as} \quad k \to \infty.
\label{DispRelAcoustic}
\end{equation}
The optical mode has the dispersion relation $\omega = \omega_o(k)$ satisfying the two asymptotic approximations
\begin{equation}
\omega_o^2(k) = \epsilon(\delta + \gamma) + \mathcal{O}(k^2) \quad \mbox{\rm as} \quad k \to 0 \eqtext{and} \omega_o^2(k) = \frac{1}{\epsilon} + \mathcal{O}(k^{-2}) \quad \mbox{\rm as} \quad k \to \infty.
\label{DispRelOptical}
\end{equation}
%As shown for example in \cite{Khusnutdinova09}, we note that 
Pure propagating solitary waves are not supported by the coupled system and are replaced with long-living radiating solitary waves \cite{KSZ, GKM}. 
%Indeed, if the wave propagates with the finite speed $v \neq 0$, then there exists a corresponding $\pm k_0 \neq 0$ such that $\omega(\pm k_0) = v k_0$, therefore, the solitary wave is in resonance with the optical mode. This applies also to the family of approximate solitary waves with the speed $v$ close to the linear phase speed
%\begin{equation}
%v_0 = \sqrt{\frac{\gamma + \delta c^2}{\gamma + \delta}}
%\end{equation}
%of the acoustic mode. However, when we use the approximation of solitary waves with the KdV or Ostrovsky equation, we do not take into account this resonance %because it occurs beyond the algebraic powers of the asymptotic expansion in $\epsilon$.

In this paper we revisit the weakly-nonlinear solution of the Cauchy problem for the coupled Boussinesq-type equations constructed in \cite{KM} with the view of extending the applicability of the solution to initial conditions with non-zero mass. We use the novel multiple-scales procedure recently devloped in \cite{KT18}. 

%%%%%%
The paper is organised as follows. In Section \ref{sec:WNL} we construct a weakly-nonlinear solution of the problem using asymptotic multiple-scales expansions for the deviations from the oscillating mean values (similarly to \cite{KT18, KMP}), using fast characteristic variables and two slow time variables \cite{KT18}. The validity of the constructed solution is examined in Section \ref{sec:Validity}, where we compare it with direct numerical simulations of the Cauchy problem. We use both the constructed weakly-nonlinear solution and direct numerical simulations to study the interaction of two radiating solitary waves in Section \ref{sec:RSW} and conclude in Section \ref{sec:Conc}. Numerical methods used in these studies are described in Appendix \ref{sec:NumMeth}.

\section{Weakly Nonlinear D'Alembert-Type Solution}
\label{sec:WNL}
Following our earlier work \cite{ KT18, KMP}, we consider the equation system \eqref{ueq} - \eqref{weq} on the periodic domain $x \in [-L, L]$ and adjust the asymptotic expansions  to the coupled system of Boussinesq-type equations. Firstly, we integrate \eqref{ueq} - \eqref{weq} in $x$ over the period $2L$ to obtain an evolution equation of the form
\begin{align}
\diffn{2}{ }{t} \int_{-L}^{L} u(x,t) \dd{x} + \epsilon \delta \int_{-L}^{L} \lb u(x,t) - w(x,t) \rb &= 0, \label{ueqmean} \\
\diffn{2}{ }{t} \int_{-L}^{L} w(x,t) \dd{x} - \epsilon \gamma \int_{-L}^{L} \lb u(x,t) - w(x,t) \rb &= 0. \label{weqmean}
\end{align}
Denoting the mean value of $u$ and $w$ as
\begin{equation}
\overline{u}(t) := \frac{1}{2L} \int_{-L}^{L} u(x,t) \dd{x}, \quad \overline{w}(t) := \frac{1}{2L} \int_{-L}^{L} w(x,t) \dd{x},
\label{MeanVal}
\end{equation}
we solve this system to describe the evolution of the mean values:
\begin{align}
\overline{u} &= d_{1} + \delta d_{2} \cos{\omega t} + d_{3} t + \delta d_{4} \sin{\omega t}, \label{uMeanVal} \\
\overline{w} &= d_{1} - \gamma d_{2} \cos{\omega t} + d_{3} t - \gamma d_{4} \sin{\omega t}. \label{wMeanVal}
\end{align}
Taking the mean value of the initial conditions \eqref{uIC}, \eqref{wIC} we obtain
\begin{equation}
d_{1} = \frac{\gamma \tilde{F}_{1} + \delta \tilde{F}_{2}}{\delta + \gamma}, \quad d_{2} = \frac{\tilde{F}_{1} - \tilde{F}_{2}}{\delta + \gamma}, \quad d_{3} = \frac{\gamma \tilde{V}_{1} + \delta \tilde{V}_{2}}{\omega \lb \delta + \gamma \rb}, \quad d_{4} = \frac{\tilde{V}_{1} - \tilde{V}_{2}}{\omega \lb \delta + \gamma \rb},
\label{MeanValdi}
\end{equation}
where we used the notation $\omega = \sqrt{\epsilon \lb \delta + \gamma \rb}$ (not to be confused with the wave frequency, which is not used any more) and
\begin{equation}
\tilde{F}_{i} = \int_{-L}^{L} F_{i}(x) \dd{x}, \quad \tilde{V}_{i} = \int_{-L}^{L} V_{i}(x) \dd{x}, \quad i = 1,2.
\label{MeanValIC}
\end{equation}
To avoid linear growth in the mean value we require that $d_{3} = 0$, corresponding to
\begin{equation}
\gamma \tilde{V}_{1} = - \delta \tilde{V}_{2}.
\label{VWCondition}
\end{equation}
In the following we consider initial conditions that satisfy the stricter condition $d_{3} = d_{4} = 0$, that is
\begin{equation}
\frac{1}{2L} \int_{-L}^{L} V_{i} \dd{x} = 0, \quad i = 1,2.
\label{UWCondition2}
\end{equation}
The latter condition appears naturally in many physical applications, and we impose it here in order to simplify our derivations. 

The calculated mean values are subtracted from $u$ and $w$ to obtain an equation with zero mean value. We take $\tilde{u} = u - \overline{u}$ and $\tilde{w} = w - \overline{w}$ to obtain the modified evolution problem for the deviations from the mean values:
\begin{align}
&\tilde{u}_{tt} - \tilde{u}_{xx} = \epsilon \lsq \frac{1}{2} \lb \tilde{u}^2 \rb_{xx} + \lb d_{1} + \delta d_{2} \cos{\omega t} \rb \tilde{u}_{xx} + \tilde{u}_{ttxx} - \delta \lb \tilde{u} - \tilde{w} \rb \rsq, \label{ueqtilde} \\
&\tilde{w}_{tt} - c^2 \tilde{w}_{xx} = \epsilon \lsq \frac{\alpha}{2} \lb \tilde{w}^2 \rb_{xx} + \alpha \lb d_{1} - \gamma d_{2} \cos{\omega t} \rb \tilde{w}_{xx} + \beta \tilde{w}_{ttxx} + \gamma \lb \tilde{u} - \tilde{w} \rb \rsq. \label{weqtilde}
\end{align}
The initial conditions take the form
\begin{align}
\color{red}
&\tilde{u}(x,0) = \tilde{F}_{1}(x) = F_{1}(x) - \overline{u}, \quad \tilde{u}_{t}(x,0) = \tilde{V}_{1}(x) = V_{1}(x), \label{uICtilde} \\
&\tilde{w}(x,0) = \tilde{F}_{2}(x) = F_{2}(x) - \overline{w}, \quad \tilde{w}_{t}(x,0) = \tilde{V}_{2}(x) = V_{2}(x), \label{wICtilde}
\end{align}
%Therefore the modified initial conditions 
and, by construction, have zero mean value. We omit the tildes in what follows.

In this paper we will consider the case when the phase speeds are close, characterised by $c^2 - 1 = \O{\epsilon}$. In this case the waves are resonant and an initial solitary wave solution in both layers will evolve into a \textit{radiating solitary wave}, that is a solitary wave with a co-propagating one-sided oscillatory tail \cite{KSZ}. Therefore we rearrange \eqref{weqtilde} so that the same characteristic variable can be used in both equations, obtaining
\begin{align}
&{u}_{tt} - {u}_{xx} = \epsilon \lsq \frac{1}{2} \lb {u}^2 \rb_{xx} + \lb d_{1} + \delta d_{2} \cos{\omega t} \rb {u}_{xx} + \beta {u}_{ttxx} - \delta \lb {u} - {w} \rb \rsq, \label{ueqtilde2} \\
&{w}_{tt} - {w}_{xx} = \epsilon \lsq \frac{\alpha}{2} \lb {w}^2 \rb_{xx} + \alpha \lb d_{1} - \gamma d_{2} \cos{\omega t} + \frac{c^2 - 1}{\epsilon} \rb {w}_{xx} + \beta {w}_{ttxx} + \gamma \lb {u} - {w} \rb \rsq,
\label{weqtilde2}
\end{align}
where we note that $(c^2 - 1)/\epsilon = \O{1}$. 
%In what follows we omit tildes for brevity. 
We look for a weakly-nonlinear solution of the form
\begin{align}
	u(x,t) =& f_{1}^{-} \lb \xi_{-}, \tau, T \rb + f_{1}^{+} \lb \xi_{+}, \tau, T \rb + \sqrt{\epsilon} P_{1} \lb \xi_{-}, \xi_{+}, \tau, T \rb + \epsilon Q_{1} \lb \xi_{-}, \xi_{+}, \tau, T \rb \notag \\
	&~+ \epsilon^{\frac{3}{2}} R_{1} \lb \xi_{-}, \xi_{+}, \tau, T \rb + \epsilon^2 S_{1} \lb \xi_{-}, \xi_{+}, \tau, T \rb + \O{\epsilon^{\frac{5}{2}}}, \label{uWNL} \\
	w(x,t) =& f_{2}^{-} \lb \xi_{-}, \tau, T \rb + f_{2}^{+} \lb \xi_{+}, \tau, T \rb + \sqrt{\epsilon} P_{2} \lb \xi_{-}, \xi_{+}, \tau, T \rb + \epsilon Q_{2} \lb \xi_{-}, \xi_{+}, \tau, T \rb \notag \\
	&~+ \epsilon^{\frac{3}{2}} R_{2} \lb \xi_{-}, \xi_{+}, \tau, T \rb + \epsilon^2 S_{2} \lb \xi_{-}, \xi_{+}, \tau, T \rb + \O{\epsilon^{\frac{5}{2}}}, \label{wWNL}
\end{align}
where we introduce fast characteristic variables and two slow time variables \cite{KT18}
\begin{equation*}
\xi_{\pm} = x \pm t, \quad \tau = \sqrt{\epsilon} t, \quad T = \epsilon t.
\end{equation*}
We aim to construct the d'Alembert-type solution on the periodic domain, similarly to \cite{KT18}. Therefore we substitute \eqref{uWNL} into \eqref{ueqtilde2} and \eqref{wWNL} into \eqref{weqtilde2} then collect terms at powers of $\sqrt{\epsilon}$ to determine expressions for all functions in the expansion. 

Here, $u$ and $w$ are $2L$-periodic functions in $x$, therefore we require that $f_{1,2}^{-}$ and $f_{1,2}^{+}$ are periodic in $\xi_{-}$ and $\xi_{+}$ respectively, and that all terms in the asymptotic expansions are products of the functions $f_{1,2}^{\pm}$ and their derivatives. Therefore all functions are periodic in $\xi_{-}, \xi_{+}$ at fixed $\xi_{+}, \xi_{-}$. Also, as the functions $f_{1,2}^{\pm}$ have zero mean i.e.
\begin{equation}
\frac{1}{2L} \int_{-L}^{L} f_{1,2}^{\pm} \dd{\xi_{\pm}} = 0,
\label{fZM}
\end{equation}
then all functions in the expansion have zero mean.

We now collect terms at increasing powers of $\sqrt{\epsilon}$. The equation is satisfied at leading order so we move on to $\O{\sqrt{\epsilon}}$. For the equation in $u$ we have
\begin{equation}
	-4 P_{1 \xi_{-} \xi_{+}} - 2 f^{-}_{1 \xi_{-} \tau} + 2 f^{+}_{1 \xi_{+} \tau} = 0.
	\label{P1eqpreav}
\end{equation}
To obtain equations for $f_{1}^{\pm}$ we average \eqref{P1eqpreav} with respect to the fast spatial variable $x$ at constant $\xi_{-}$ and $\xi_{+}$ (see \cite{KM, KT18}). Let us first consider the averaging of $P_{1}$. At constant $\xi_{-}$ we have
\begin{equation}
	\frac{1}{2L} \int_{-L}^{L} P_{1 \xi_{-} \xi_{+}} \dd{x} = \frac{1}{4L} \int_{-2L - \xi_{-}}^{2L - \xi_{-}} P_{1 \xi_{-} \xi_{+}} \dd{\xi_{+}} = \frac{1}{4L} \lsq P_{1 \xi_{-}} \rsq_{-2L - \xi_{-}}^{2L - \xi_{-}} = 0,
	\label{Pav}
\end{equation}
and similarly for averaging at constant $\xi_{+}$. Therefore under the averaging we have $P_{1 \xi_{-} \xi_{+}} = 0$. Applying the averaging to \eqref{P1eqpreav} we have two equations:
\begin{equation}
	f^{-}_{1 \xi_{-} \tau} = 0 \eqtext{and} f^{+}_{1 \xi_{+} \tau} = 0,
	\label{f1eqdiff}
\end{equation}
which implies that
\begin{equation}
	f^{-}_{1} = \tilde{f}^{-}_{1} \lb \xi_{-}, T \rb + B^{-} \lb \tau, T \rb \eqtext{and} f^{+}_{1} = \tilde{f}^{+}_{1} \lb \xi_{+}, T \rb + B^{+} \lb \tau, T \rb.
	\label{f1eq}
\end{equation}
As we have zero mean of all functions in the expansion, we have $B^{\pm} = 0$. Similarly for the equation in $w$,
\begin{equation}
	-4 P_{2 \xi_{-} \xi_{+}} - 2 f^{-}_{2 \xi_{-} \tau} + 2 f^{+}_{2 \xi_{+} \tau} = 0,
	\label{P2eqpreav}
\end{equation}
which after averaging at constant $\xi_{-}$ or $\xi_{+}$ and applying the same reasoning as for $f_{1}^{\pm}$, we obtain
\begin{equation}
	f^{-}_{2} = \tilde{f}^{-}_{2} \lb \xi_{-}, T \rb \eqtext{and} f^{+}_{2} = \tilde{f}^{+}_{2} \lb \xi_{+}, T \rb.
	\label{f2eq}
\end{equation}
Substituting \eqref{f1eq} into \eqref{P1eqpreav} gives
\begin{equation}
	P_{1 \xi_{-} \xi_{+}} = 0 \RA P_{1} = g_{1}^{-} \lb \xi_{-}, \tau, T \rb + g_{1}^{+} \lb \xi_{+}, \tau, T \rb.
	\label{P1eq}
\end{equation}
Similarly substituting \eqref{f2eq} into \eqref{P2eqpreav} gives
\begin{equation}
	P_{2 \xi_{-} \xi_{+}} = 0 \RA P_{2} = g_{2}^{-} \lb \xi_{-}, \tau, T \rb + g_{2}^{+} \lb \xi_{+}, \tau, T \rb.
	\label{P2eq}
\end{equation}
We omit the tildes on $f_{1,2}^{\pm}$ in subsequent steps. The initial condition for $f_{1,2}^{\pm}$ is found by substituting \eqref{uWNL} into \eqref{uICtilde} and \eqref{wWNL} into \eqref{wICtilde} and comparing terms at $\O{1}$ to obtain d'Alembert-type formulae for $f_{1,2}^{\pm}$ of the form
\begin{equation}
	\lset
	\begin{aligned}
		\left. f_{1}^{-} + f_{1}^{+} \right |_{T = 0} &= \tilde{F}_{1}(x), \\
		\left. -f_{1 \xi_{-}}^{-} + f_{1 \xi_{+}}^{+} \right |_{T = 0} &= \tilde{V}_{1}(x),
	\end{aligned}
	\right.
	\RA
	f_{1}^{\pm}|_{T = 0} = \frac{1}{2} \lb \tilde{F}_{1} \lb x \pm t \rb \pm \int_{-L}^{x \pm t} \tilde{V}_{1} \lb \sigma \rb \dd{\sigma} \rb,
	\label{f1IC}
\end{equation}
and
\begin{equation}
	\lset
	\begin{aligned}
		\left. f_{2}^{-} + f_{2}^{+} \right |_{T = 0} &= \tilde{F}_{2}(x), \\
		\left. -f_{2 \xi_{-}}^{-} + f_{2 \xi_{+}}^{+} \right |_{T = 0} &= \tilde{V}_{2}(x),
	\end{aligned}
	\right.
	\RA
	f_{2}^{\pm}|_{T = 0} = \frac{1}{2} \lb \tilde{F}_{2} \lb x \pm t \rb \pm \int_{-L}^{x \pm t} \tilde{V}_{2} \lb \sigma \rb \dd{\sigma} \rb.
	\label{f2IC}
\end{equation}
We now move on to the terms at $\O{\epsilon}$, using the results from the previous order. For the equation governing $u$ we obtain
\begin{align}
-4Q_{1 \xi_{-} \xi_{+}} &= 2 g_{1 \xi_{-} \tau}^{-} + \lb 2 f_{1 T}^{-} + f_{1}^{-} f_{1 \xi_{-}}^{-} + d_{1} f_{1 \xi_{-}}^{-} + f_{1 \xi_{-} \xi_{-} \xi_{-}}^{-} \rb_{\xi_{-}} - \delta \lb f_{1}^{-} - f_{2}^{-} \rb \notag \\
&~~~- 2 g_{1 \xi_{+} \tau}^{+} + \lb -2 f_{1 T}^{+} + f_{1}^{+} f_{1 \xi_{+}}^{+} + d_{1} f_{1 \xi_{+}}^{+} + f_{1 \xi_{+} \xi_{+} \xi_{+}}^{+} \rb_{\xi_{+}} - \delta \lb f_{1}^{+} - f_{2}^{+} \rb \notag \\
&~~~+ d_{2} \delta \cos{\lb \tilde{\omega} t \rb} \lb f_{1 \xi_{-} \xi_{-}}^{-} + f_{1 \xi_{+} \xi_{+}}^{+} \rb + f_{1 \xi_{-} \xi_{-}}^{-} f_{1}^{+} + 2 f_{1 \xi_{-}}^{-} f_{1 \xi_{+}}^{+} + f_{1}^{-} f_{1 \xi_{+} \xi_{+}}^{+},
\label{Q1eqpreav}
\end{align}
where we have introduced the notation $\tilde{\omega} = \sqrt{\delta + \gamma}$. Similarly for the equation governing $w$,
\begin{align}
-4Q_{2 \xi_{-} \xi_{+}} &= 2 g_{2 \xi_{-} \tau}^{-} + \lb 2 f_{2 T}^{-} + \alpha f_{2}^{-} f_{2 \xi_{-}}^{-} + \alpha d_{1} f_{2 \xi_{-}}^{-} + \beta f_{2 \xi_{-} \xi_{-} \xi_{-}}^{-} \rb_{\xi_{-}} + \gamma \lb f_{1}^{-} - f_{2}^{-} \rb \notag \\
&~~~- 2 g_{2 \xi_{+} \tau}^{+} + \lb -2 f_{2 T}^{+} + \alpha f_{2}^{+} f_{2 \xi_{+}}^{+} + \alpha d_{1} f_{2 \xi_{+}}^{+} + \beta f_{2 \xi_{+} \xi_{+} \xi_{+}}^{+} \rb_{\xi_{+}} + \gamma \lb f_{1}^{+} - f_{2}^{+} \rb \notag \\
&~~~+ \lb \frac{c^2 - 1}{\epsilon} - \alpha d_{2} \gamma \cos{\lb \tilde{\omega} t \rb} \rb \lb f_{2 \xi_{-} \xi_{-}}^{-} + f_{2 \xi_{+} \xi_{+}}^{+} \rb \notag \\
&~~~+ \alpha \lb f_{2 \xi_{-} \xi_{-}}^{-} f_{2}^{+} + 2 f_{2 \xi_{-}}^{-} f_{2 \xi_{+}}^{+} + f_{2}^{-} f_{2 \xi_{+} \xi_{+}}^{+} \rb.
\label{Q2eqpreav}
\end{align}
Averaging \eqref{Q1eqpreav} and \eqref{Q2eqpreav} at constant $\xi_{-}$ or constant $\xi_{+}$ gives the system of equations
\begin{align}
\pm 2 g_{1 \xi_{\pm} \tau}^{\pm} &= d_{2} \delta \cos{\lb \tilde{\omega} t \rb} f_{1 \xi_{\pm} \xi_{\pm}}^{\pm} + A_{1}^{\pm} \lb \xi_{\pm}, T \rb, \label{g1Sec} \\
\pm 2 g_{2 \xi_{\pm} \tau}^{\pm} &= - \alpha d_{2} \gamma \cos{\lb \tilde{\omega} t \rb} f_{2 \xi_{\pm} \xi_{\pm}}^{\pm} + A_{2}^{\pm} \lb \xi_{\pm}, T \rb, \label{g2Sec}
\end{align}
where
\begin{equation}
	A_{1}^{\pm} = \lb \mp 2 f_{1 T}^{\pm} + f_{1}^{\pm} f_{1 \xi_{\pm}}^{\pm} + d_{1} f_{1 \xi_{\pm}}^{\pm} + f_{1 \xi_{\pm} \xi_{\pm} \xi_{\pm}}^{\pm} \rb_{\xi_{\pm}} - \delta \lb f_{1}^{\pm} - f_{2}^{\pm} \rb,
	\label{A1}
\end{equation}
and
\begin{equation}
	A_{2}^{\pm} = \lb \mp 2 f_{2 T}^{\pm} + \alpha f_{2}^{\pm} f_{2 \xi_{\pm}}^{\pm} + \lb \alpha d_{1} + \frac{c^2 - 1}{\epsilon} \rb f_{2 \xi_{\pm}}^{\pm} + \beta f_{2 \xi_{\pm} \xi_{\pm} \xi_{\pm}}^{\pm} \rb_{\xi_{\pm}} + \gamma \lb f_{1}^{\pm} - f_{2}^{\pm} \rb.
	\label{A2}
\end{equation}
To avoid secular terms we require that $A_{1} = 0$ and $A_{2} = 0$. Therefore we get a system of coupled Ostrovsky equations for $f_{1,2}^{\pm}$, of the form
\begin{align}
	&\lb \mp 2 f_{1 T}^{\pm} + f_{1}^{\pm} f_{1 \xi_{\pm}}^{\pm} + d_{1} f_{1 \xi_{\pm}}^{\pm} + f_{1 \xi_{\pm} \xi_{\pm} \xi_{\pm}}^{\pm} \rb_{\xi_{\pm}} = \delta \lb f_{1}^{\pm} - f_{2}^{\pm} \rb, \notag \\
	&\lb \mp 2 f_{2 T}^{\pm} + \alpha f_{2}^{\pm} f_{2 \xi_{\pm}}^{\pm} + \lb \alpha d_{1} + \frac{c^2 - 1}{\epsilon} \rb f_{2 \xi_{\pm}}^{\pm} + \beta f_{2 \xi_{\pm} \xi_{\pm} \xi_{\pm}}^{\pm} \rb_{\xi_{\pm}} = \gamma \lb f_{2}^{\pm} - f_{1}^{\pm} \rb.
	\label{fcOst}
\end{align}
Integrating \eqref{g1Sec} and \eqref{g2Sec} with these conditions applied we can find an equation for $g_{1,2}^{\pm}$ of the form
\begin{equation}
	g_{1}^{\pm} = \pm \frac{d_{2} \delta}{2 \tilde{\omega}} \sin{\lb \tilde{\omega} \tau \rb} f_{1 \xi_{\pm}}^{\pm} + G_{1}^{\pm} \lb \xi_{\pm}, T \rb = \pm \theta_{1} f_{1 \xi_{\pm}}^{\pm} + G_{1}^{\pm} \lb \xi_{\pm}, T \rb,
	\label{g1eq}
\end{equation}
and
\begin{equation}
	g_{2}^{\pm} = \mp \frac{\alpha d_{2} \gamma}{2 \tilde{\omega}} \sin{\lb \tilde{\omega} \tau \rb} f_{2 \xi_{\pm}}^{\pm} + G_{2}^{\pm} \lb \xi_{\pm}, T \rb = \mp \theta_{2} f_{2 \xi_{\pm}}^{\pm} + G_{2}^{\pm} \lb \xi_{\pm}, T \rb,
	\label{g2eq}
\end{equation}
where $G_{1}^{\pm}$ and $G_{2}^{\pm}$ are functions to be found and we introduce
\begin{equation}
	\theta_{1} = \frac{d_{2} \delta}{2 \tilde{\omega}} \sin{\lb \tilde{\omega} \tau \rb}, \quad \theta_{2} = \frac{\alpha d_{2} \gamma}{2 \tilde{\omega}} \sin{\lb \tilde{\omega} \tau \rb},
	\label{theta12}
\end{equation}
for convenience. Substituting \eqref{fcOst} and \eqref{g1eq} into \eqref{Q1eqpreav} and integrating gives
\begin{equation}
	Q_{1} = h_{1}^{-} \lb \xi_{-}, \tau, T \rb + h_{1}^{+} \lb \xi_{+}, \tau, T \rb + h_{1c} \lb \xi_{-}, \xi_{+}, T \rb,
	\label{Q1eq}
\end{equation}
where
\begin{equation}
h_{1c} = - \frac{1}{4} \lb f_{1 \xi_{-}} \int_{-L}^{\xi_{+}} f_{1}^{+} \lb s \rb \dd{s} + 2 f_{1}^{-} f_{1}^{+} + f_{1 \xi_{+}}^{+} \int_{-L}^{\xi_{-}} f_{1}^{-} \lb s \rb \dd{s} \rb.
\label{h1c}
\end{equation}
Similarly substituting \eqref{fcOst} and \eqref{g2eq} into \eqref{Q2eqpreav} and integrating gives
\begin{equation}
	Q_{2} = h_{2}^{-} \lb \xi_{-}, \tau, T \rb + h_{2}^{+} \lb \xi_{+}, \tau, T \rb + h_{2c} \lb \xi_{-}, \xi_{+}, T \rb,
	\label{Q2eq}
\end{equation}
where
\begin{equation}
h_{2c} = - \frac{\alpha}{4} \lb f_{2 \xi_{-}} \int_{-L}^{\xi_{+}} f_{2}^{+} \lb s \rb \dd{s} + 2 f_{2}^{-} f_{2}^{+} + f_{2 \xi_{+}}^{+} \int_{-L}^{\xi_{-}} f_{2}^{-} \lb s \rb \dd{s} \rb.
\label{h2c}
\end{equation}
The initial condition for $G_{1,2}^{\pm}$ is found by substituting \eqref{uWNL} and \eqref{wWNL} into \eqref{uICtilde} and \eqref{wICtilde} respectively, and comparing terms at $\O{\sqrt{\epsilon}}$, taking account of the results found in \eqref{g1eq} and \eqref{g2eq}. Therefore we obtain
\begin{equation}
	\lset
	\begin{aligned}
		\left. \theta_{1} f_{1 \xi_{-}}^{-} + \theta_{1} f_{1 \xi_{+}}^{+} + G_{1}^{-} + G_{1}^{+} \right |_{T = 0} &= 0, \\
		\left. -\theta_{1} f_{1 \xi_{-} \xi_{-}}^{-} + \theta_{1} f_{1 \xi_{+} \xi_{+}}^{+} - G_{1 \xi_{-}}^{-} + G_{1 \xi_{+}}^{+} \right |_{T = 0} &= 0,
	\end{aligned}
	\right.
	\RA G_{1}^{\pm} = 0,
	\label{G1IC}
\end{equation}
as we see from \eqref{theta12} that $\theta_{1}|_{T=0} = 0$. Similarly we have 
\begin{equation}
	\lset
	\begin{aligned}
		\left. \theta_{2} f_{2 \xi_{-}}^{-} + \theta_{2} f_{2 \xi_{+}}^{+} + G_{2}^{-} + G_{2}^{+} \right |_{T = 0} &= 0, \\
		\left. -\theta_{2} f_{2 \xi_{-} \xi_{-}}^{-} + \theta_{2} f_{2 \xi_{+} \xi_{+}}^{+} - G_{2 \xi_{-}}^{-} + G_{2 \xi_{+}}^{+} \right |_{T = 0} &= 0,
	\end{aligned}
	\right.
	\RA G_{2}^{\pm} = 0.
	\label{G2IC}
\end{equation}
We now consider terms at $\O{\epsilon^{3/2}}$. Substituting the results obtained at previous orders of $\epsilon$ into the weakly-nonlinear expansion \eqref{uWNL} and \eqref{wWNL}, then substituting this into \eqref{ueqtilde2} and \eqref{weqtilde2} and gathering terms at $\O{\epsilon^{3/2}}$ gives
\begin{align}
-4R_{1 \xi_{-} \xi_{+}} &= 2 h_{1 \xi_{-} \tau}^{-} - 2 h_{1 \xi_{+} \tau} + \lb 2 g_{1 T}^{-} + \lb f_{1}^{-} g_{1}^{-} \rb_{\xi_{-}} + d_{1} g_{1 \xi_{-}}^{-} + g_{1 \xi_{-} \xi_{-} \xi_{-}}^{-} \rb_{\xi_{-}}  - \delta \lb g_{1}^{-} - g_{2}^{-} \rb \notag \\
&~~~- g_{1 \tau \tau}^{-} - g_{1 \tau \tau}^{+} + \lb - 2 g_{1 T}^{+} + \lb f_{1}^{+} g_{1}^{+} \rb_{\xi_{+}} + d_{1} g_{1 \xi_{+}}^{+} + g_{1 \xi_{+} \xi_{+} \xi_{+}}^{+} \rb_{\xi_{+}} - \delta \lb g_{1}^{+} - g_{2}^{+} \rb \notag \\
&~~~+ d_{2} \delta \cos{\lb \tilde{\omega} \tau \rb} \lb g_{1 \xi_{-} \xi_{-}}^{-} + g_{1 \xi_{+} \xi_{+}}^{+} \rb + g_{1 \xi_{-} \xi_{-}}^{-} f_{1}^{+} + 2 g_{1 \xi_{-}}^{-} f_{1 \xi_{+}}^{+} + g_{1}^{-} f_{1 \xi_{+} \xi_{+}}^{+} \notag \\
&~~~+ g_{1 \xi_{+} \xi_{+}}^{+} f_{1}^{-} + 2 g_{1 \xi_{+}}^{+} f_{1 \xi_{-}}^{-} + g_{1}^{+} f_{1 \xi_{-} \xi_{-}}^{-},
\label{R1eqpreav}
\end{align}
and
\begin{align}
-4R_{2 \xi_{-} \xi_{+}} &= 2 h_{2 \xi_{-} \tau}^{-} + \lb 2 g_{2 T}^{-} + \alpha \lb f_{2}^{-} g_{2}^{-} \rb_{\xi_{-}} + \alpha d_{1} g_{2 \xi_{-}}^{-} + \beta g_{2 \xi_{-} \xi_{-} \xi_{-}}^{-} \rb_{\xi_{-}} + \gamma \lb g_{1}^{-} - g_{2}^{-} \rb \notag \\
&~~~- 2 h_{2 \xi_{+} \tau}^{+} + \lb - 2 g_{2 T}^{+} + \alpha \lb f_{2}^{+} g_{2}^{+} \rb_{\xi_{+}} + \alpha d_{1} g_{2 \xi_{+}}^{+} + \beta g_{2 \xi_{+} \xi_{+} \xi_{+}}^{+} \rb_{\xi_{+}} + \gamma \lb g_{1}^{+} - g_{2}^{+} \rb \notag \\
&~~~+ \lb \frac{c^2 - 1}{\epsilon} - \alpha d_{2} \gamma \cos{\lb \tilde{\omega} \tau \rb} \rb \lb g_{2 \xi_{-} \xi_{-}}^{-} + g_{2 \xi_{+} \xi_{+}}^{+} \rb - g_{2 \tau \tau}^{-} - g_{2 \tau \tau}^{+} \notag \\
&~~~+ \alpha \lsq g_{2 \xi_{-} \xi_{-}}^{-} f_{2}^{+} + 2 g_{2 \xi_{-}}^{-} f_{2 \xi_{+}}^{+} + g_{2}^{-} f_{2 \xi_{+} \xi_{+}}^{+} + g_{2 \xi_{+} \xi_{+}}^{+} f_{2}^{-} + 2 g_{2 \xi_{+}}^{+} f_{2 \xi_{-}}^{-} + g_{2}^{+} f_{2 \xi_{-} \xi_{-}}^{-} \rsq.
\label{R2eqpreav}
\end{align}
Substituting \eqref{g1eq} into \eqref{R1eqpreav} and averaging at constant $\xi_{-}$ or constant $\xi_{+}$ gives
\begin{align}
\pm 2 h_{1 \xi_{\pm} \tau}^{\pm} &= \pm \theta_{1} \lb \mp 2 f_{1 T}^{\pm} + f_{1}^{\pm} f_{1 \xi_{\pm}}^{\pm} + d_{1} f_{1 \xi_{\pm}}^{\pm} + f_{1 \xi_{\pm} \xi_{\pm} \xi_{\pm}} \rb_{\xi_{\pm} \xi_{\pm}} \mp \delta \lb \theta_{1} f_{1}^{\pm} + \theta_{2} f_{2}^{\pm} \rb_{\xi_{\pm}} \notag \\
&~~~+ \lb \mp 2 G_{1 T}^{\pm} + \lb f_{1}^{\pm} G_{1}^{\pm} \rb_{\xi_{\pm}} + d_{1} G_{1 \xi_{\pm}}^{\pm} + G_{1 \xi_{\pm} \xi_{\pm} \xi_{\pm}} \rb_{\xi_{\pm}} - \delta \lb G_{1}^{\pm} - G_{2}^{\pm} \rb \notag \\
&~~~\pm \theta_{1} \tilde{\omega}^2 f_{1 \xi_{\pm}}^{\pm} \pm d_{2} \delta \cos{\lb \tilde{\omega} \tau \rb} \theta_{1} f_{1 \xi_{\pm} \xi_{\pm} \xi_{\pm}}^{\pm}. 
\label{R1averaged}
\end{align}
If we differentiate the equation for $f_{1}$ in \eqref{fcOst} with respect to the appropriate characteristic variable, we can eliminate some terms in the first line in \eqref{R1averaged} to obtain an expression for $h_{1 \xi_{\pm} \tau}^{\pm}$ of the form
\begin{equation}
	2 h_{1 \xi_{\pm} \tau}^{\pm} = \theta_{1} \tilde{\omega}^2 f_{1 \xi_{\pm}}^{\pm} - \delta \lb \theta_{1} + \theta_{2} \rb f_{2 \xi_{\pm}}^{\pm} + \theta_{1} d_{2} \delta \cos{\lb \tilde{\omega} \tau \rb} f_{1 \xi_{\pm} \xi_{\pm} \xi_{\pm}}^{\pm} + \tilde{G}_{1}^{\pm} \lb \xi_{\pm}, T \rb,
	\label{h1diff}
\end{equation}
where
\begin{equation}
	\tilde{G}_{1}^{\pm} \lb \xi_{\pm}, T \rb = \lb \mp 2 G_{1 T}^{\pm} + \lb f_{1}^{\pm} G_{1}^{\pm} \rb_{\xi_{\pm}} + d_{1} G_{1 \xi_{\pm}}^{\pm} + G_{1 \xi_{\pm} \xi_{\pm} \xi_{\pm}}^{\pm} \rb_{\xi_{\pm}} - \delta \lb G_{1}^{\pm} - G_{2}^{\pm} \rb.
	\label{G1tildeeq}
\end{equation}
To avoid secular terms again we require that $\tilde{G}_{1}^{\pm} = 0$ and therefore we have an equation for $G_{1}^{\pm}$ of the form
\begin{equation}
	\lb \mp 2 G_{1 T}^{\pm} + \lb f_{1}^{\pm} G_{1}^{\pm} \rb_{\xi_{\pm}} + d_{1} G_{1 \xi_{\pm}}^{\pm} + G_{1 \xi_{\pm} \xi_{\pm} \xi_{\pm}}^{\pm} \rb_{\xi_{\pm}} = \delta \lb G_{1}^{\pm} - G_{2}^{\pm} \rb.
	\label{G1eq}
\end{equation}
Integrating \eqref{h1diff} we obtain
\begin{equation}
	h_{1}^{\pm} = -\frac{\delta d_{2}}{4} \cos{\lb \omega t \rb} f_{1}^{\pm} + \frac{\delta d_{2} \lb \delta + \alpha \gamma \rb}{4 \tilde{\omega}^2} \cos{\lb \omega t \rb} f_{2}^{\pm} - \frac{\delta^2 d_{2}^2}{8 \tilde{\omega}^2} \cos^2 \lb \omega t \rb f_{1 \xi_{\pm} \xi_{\pm}}^{\pm} + \phi_{1}^{\pm} \lb \xi_{\pm}, T \rb,
	\label{h1eq}
\end{equation}
where the function $\phi_{1}^{\pm}$ is to be found. Averaging \eqref{R2eqpreav} at constant $\xi_{-}$ or $\xi_{+}$, and using \eqref{g2eq} and \eqref{fcOst} as was done above, we get an expression for $h_{2}$ of the form
\begin{equation}
	2 h_{2 \xi_{\pm} \tau}^{\pm} = -\theta_{2} \tilde{\omega}^2 f_{2 \xi_{\pm}}^{\pm} + \gamma \lb \theta_{1} + \theta_{2} \rb f_{1 \xi_{\pm}}^{\pm} + \theta_{2} d_{2} \gamma \cos{\lb \tilde{\omega} \tau \rb} f_{2 \xi_{\pm} \xi_{\pm} \xi_{\pm}}^{\pm} + \tilde{G}_{2}^{\pm} \lb \xi_{\pm}, T \rb,
	\label{h2diff}
\end{equation}
where
\begin{align}
	\tilde{G}_{2}^{\pm} \lb \xi_{\pm}, T \rb &= \lb \mp 2 G_{2 T}^{\pm} + \alpha \lb f_{2}^{\pm} G_{2}^{\pm} \rb_{\xi_{\pm}} + \alpha d_{1} G_{2 \xi_{\pm}}^{\pm} + \frac{c^2 - 1}{\epsilon} G_{2 \xi_{\pm}}^{\pm} + \beta G_{2 \xi_{\pm} \xi_{\pm} \xi_{\pm}}^{\pm} \rb_{\xi_{\pm}} \notag \\
	&~~~+ \gamma \lb G_{1}^{\pm} - G_{2}^{\pm} \rb.
	\label{G2tildeeq}
\end{align}
As before we require that $\tilde{G}_{2}^{\pm} = 0$ and therefore we have an equation for $G_{2}^{\pm}$ of the form
\begin{equation}
	\lb \mp 2 G_{2 T}^{\pm} + \alpha \lb f_{2}^{\pm} G_{2}^{\pm} \rb_{\xi_{\pm}} + \lb \alpha d_{1} + \frac{c^2 - 1}{\epsilon} \rb G_{2 \xi_{\pm}}^{\pm} + \beta G_{2 \xi_{\pm} \xi_{\pm} \xi_{\pm}}^{\pm} \rb_{\xi_{\pm}} = \gamma \lb G_{2}^{\pm} - G_{1}^{\pm} \rb.
	\label{G2eq}
\end{equation}
Taking into account the initial condition in \eqref{G1IC}, \eqref{G2IC}, and the form of \eqref{G1eq} and \eqref{G2eq} we see that $G_{1,2}^{\pm} \equiv 0$. Integrating \eqref{h2diff} gives
\begin{equation}
	h_{2}^{\pm} = \frac{\alpha \gamma d_{2}}{4} \cos{\lb \omega t \rb} f_{2}^{\pm} - \frac{\gamma d_{2} \lb \delta + \alpha \gamma \rb}{4 \tilde{\omega}^2} \cos{\lb \omega t \rb} f_{1}^{\pm} - \frac{\alpha^2 \gamma^2 d_{2}^2}{8 \tilde{\omega}^2} \cos^2 \lb \omega t \rb f_{2 \xi_{\pm} \xi_{\pm}}^{\pm} + \phi_{2}^{\pm} \lb \xi_{\pm}, T \rb,
	\label{h2eq}
\end{equation}
where again we need to find the function $\phi_{2}^{\pm}$. Substituting \eqref{h1eq} into \eqref{R1eqpreav} and integrating with respect to the appropriate characteristic variables gives
\begin{equation}
	R_{1} = \psi_{1}^{-} \lb \xi_{-}, \tau, T \rb + \psi_{1}^{+} \lb \xi_{+}, \tau, T \rb + \psi_{1c} \lb \xi_{-}, \xi_{+}, T \rb,
	\label{R1eq}
\end{equation}
where
\begin{equation}
	\psi_{1c} = -\frac{\theta_{1}}{4} \lsq - f_{1 \xi_{-} \xi_{-}}^{-} \int_{-L}^{\xi_{+}} f_{1}^{+} \lb s \rb \dd{s} - f_{1 \xi_{-}}^{-} f_{1}^{+} + f_{1}^{-} f_{1 \xi_{+}}^{+} + f_{1 \xi_{+} \xi_{+}}^{+} \int_{-L}^{\xi_{-}} f_{1}^{-} \lb s \rb \dd{s} \rsq.
	\label{psi1c}
\end{equation}
In a similar way we find an expression for $R_{2}$ by substituting \eqref{h2eq} into \eqref{R2eqpreav} and integrating with respect to the appropriate characteristic variables to obtain
\begin{equation}
	R_{2} = \psi_{2}^{-} \lb \xi_{-}, \tau, T \rb + \psi_{2}^{+} \lb \xi_{+}, \tau, T \rb + \psi_{2c} \lb \xi_{-}, \xi_{+}, T \rb,
	\label{R2eq}
\end{equation}
where
\begin{equation}
	\psi_{2c} = -\frac{\alpha \theta_{2}}{4} \lsq f_{2 \xi_{-} \xi_{-}}^{-} \int_{-L}^{\xi_{+}} f_{2}^{+} \lb s \rb \dd{s} + f_{2 \xi_{-}}^{-} f_{2}^{+} - f_{2}^{-} f_{2 \xi_{+}}^{+} - f_{2 \xi_{+} \xi_{+}}^{+} \int_{-L}^{\xi_{-}} f_{2}^{-} \lb s \rb \dd{s} \rsq.
	\label{psi2c}
\end{equation}
The initial condition for the function $\phi_{1}^{\pm}$ is found by again substituting \eqref{uWNL} into \eqref{uICtilde} and comparing terms at $\O{\epsilon}$, taking account of \eqref{h1eq}. Therefore we obtain for $\phi_{1}^{\pm}$
\begin{align}
	&\lset
	\begin{aligned}
		\left. h_{1}^{-} + h_{1}^{+} + h_{1c} \right |_{T = 0}  &= 0, \\
		\left. f_{1 T}^{-} + f_{1 T}^{+} + g_{1 \tau}^{-} + g_{1 \tau}^{+} - h_{1 \xi_{-}}^{-} + h_{1 \xi_{+}}^{+} - h_{1c \xi_{-}} + h_{1c \xi_{+}} \right |_{T = 0} &= 0,
	\end{aligned}
	\right.
	\notag \\
	&\RA \phi_{1}^{\pm} = \frac{1}{2} \lb J_{1} \mp \int_{-L}^{\xi_{\pm}} K_{1} \lb s \rb \dd{s} \rb,
	\label{phi1IC}
\end{align}
where
\begin{align}
	J_{1} &= \frac{\delta d_{2}}{4} \lb f_{1}^{-} + f_{1}^{+} \rb - \frac{\delta d_{2} \lb \delta + \alpha \gamma \rb}{4 \tilde{\omega}^2} \lb f_{2}^{-} + f_{2}^{+} \rb + \frac{\delta^2 d_{2}^2}{8 \tilde{\omega}^2} \lb f_{1 \xi_{-} \xi_{-}}^{-} + f_{1 \xi_{+} \xi_{+}}^{+} \rb - 2 h_{1c}, \notag \\
	K_{1} &= f_{1 T}^{-} + f_{1 T}^{+} + g_{1 \tau}^{-} + g_{1 \tau}^{+}.
	\label{JK1}
\end{align}
Similarly for $\phi_{2}^{\pm}$ we find the initial condition by substituting \eqref{wWNL} into \eqref{wICtilde} and comparing terms at $\O{\epsilon}$, using \eqref{h2eq}. This gives
\begin{align}
	&\lset
	\begin{aligned}
		\left. h_{2}^{-} + h_{2}^{+} + h_{2c} \right |_{T = 0}  &= 0, \\
		\left. f_{2 T}^{-} + f_{2 T}^{+} + g_{2 \tau}^{-} + g_{2 \tau}^{+} - h_{2 \xi_{-}}^{-} + h_{2 \xi_{+}}^{+} - h_{2c \xi_{-}} + h_{2c \xi_{+}} \right |_{T = 0} &= 0,
	\end{aligned}
	\right.
	\notag \\
	&\RA \phi_{2}^{\pm} = \frac{1}{2} \lb J_{2} \mp \int_{-L}^{\xi_{\pm}} K_{2} \lb s \rb \dd{s} \rb,
	\label{phi2IC}
\end{align}
where
\begin{align}
	J_{2} &= -\frac{\alpha \gamma d_{2}}{4} \lb f_{2}^{-} + f_{2}^{+} \rb + \frac{\gamma d_{2} \lb \delta + \alpha \gamma \rb}{4 \tilde{\omega}^2} \lb f_{1}^{-} + f_{1}^{+} \rb + \frac{\alpha^2 \gamma^2 d_{2}^2}{8 \tilde{\omega}^2} \lb f_{2 \xi_{-} \xi_{-}}^{-} + f_{2 \xi_{+} \xi_{+}}^{+} \rb - 2 h_{2c}, \notag \\
	K_{2} &= f_{2 T}^{-} + f_{2 T}^{+} + g_{2 \tau}^{-} + g_{2 \tau}^{+}.
	\label{JK2}
\end{align}
To find an equation governing $\phi_{1,2}^{\pm}$ we need to retain terms at $\O{\epsilon^2}$ in the original expansion. All coupling terms in the expansion are gathered in one function for convenience as we do not require them to determine $\phi_{1,2}^{\pm}$. Gathering terms at $\O{\epsilon^2}$ we have
\begin{align}
- 4 S_{1 \xi_{-} \xi_{+}} &= -f_{1 T T}^{-} - f_{1 T T}^{+} - 2 g_{1 \tau T}^{-} - 2 g_{1 \tau T}^{+} - h_{1 \tau \tau}^{-} - h_{1 \tau \tau}^{+} + 2 h_{1 \xi_{-} T}^{-} - 2 h_{1 \xi_{+} T}^{+} \notag \\
&~~~+ 2 \psi_{1 \xi_{-} \tau}^{-} - 2 \psi_{1 \xi_{+} \tau}^{+} + \lb f_{1}^{-} h_{1}^{-} \rb_{\xi_{-} \xi_{-}} + \lb f_{1}^{+} h_{1}^{+} \rb_{\xi_{+} \xi_{+}} + \frac{1}{2} \lb g_{1}^{-^{2}} \rb_{\xi_{-} \xi_{-}} + \frac{1}{2} \lb g_{1}^{+^{2}} \rb_{\xi_{+} \xi_{+}} \notag \\
&~~~+ d_{1} h_{1 \xi_{-} \xi_{-}}^{-} + d_{1} h_{1 \xi_{+} \xi_{+}}^{+} + d_{2} \delta \cos{\lb \tilde{\omega} \tau \rb} \lb h_{1 \xi_{-} \xi_{-}}^{-} + h_{1 \xi_{+} \xi_{+}}^{+} \rb + h_{1 \xi_{-} \xi_{-} \xi_{-} \xi_{-}}^{-} \notag \\
&~~~+ h_{1 \xi_{+} \xi_{+} \xi_{+} \xi_{+}}^{+} - 2 g_{1 \xi_{-} \xi_{-} \xi_{-} \tau}^{-} + 2 g_{1 \xi_{+} \xi_{+} \xi_{+} \tau}^{+} - 2f_{1 \xi_{-} \xi_{-} \xi_{-} T}^{-} + 2 f_{1 \xi_{+} \xi_{+} \xi_{+} T}^{+} \notag \\
&~~~- \delta \lb h_{1}^{-} - h_{2}^{-} + h_{1}^{+} - h_{2}^{+} \rb - 4 \mu_{1c},
\label{S1eqpreav}
\end{align}
where $\mu_{1c}$ is the coupling terms at this order, and
\begin{align}
- 4 S_{2 \xi_{-} \xi_{+}} &= -f_{2 T T}^{-} - f_{2 T T}^{+} - 2 g_{2 \tau T}^{-} - 2 g_{2 \tau T}^{+} - h_{2 \tau \tau}^{-} - h_{2 \tau \tau}^{+} + 2 h_{2 \xi_{-} T}^{-} - 2 h_{2 \xi_{+} T}^{+} + 2 \psi_{2 \xi_{-} \tau}^{-} \notag \\
&~~~- 2 \psi_{2 \xi_{+} \tau}^{+} + \alpha \lb f_{2}^{-} h_{2}^{-} \rb_{\xi_{-} \xi_{-}} + \alpha \lb f_{2}^{+} h_{2}^{+} \rb_{\xi_{+} \xi_{+}} + \frac{\alpha }{2} \lb g_{2}^{-^{2}} \rb_{\xi_{-} \xi_{-}} + \frac{\alpha}{2} \lb g_{2}^{+^{2}} \rb_{\xi_{+} \xi_{+}} \notag \\
&~~~+ \alpha d_{1} h_{2 \xi_{-} \xi_{-}}^{-} + \alpha d_{1} h_{2 \xi_{+} \xi_{+}}^{+} - \alpha d_{2} \gamma \cos{\lb \tilde{\omega} \tau \rb} \lb h_{2 \xi_{-} \xi_{-}}^{-} + h_{2 \xi_{+} \xi_{+}}^{+} \rb + \beta h_{2 \xi_{-} \xi_{-} \xi_{-} \xi_{-}}^{-} \notag \\
&~~~+ \beta h_{2 \xi_{+} \xi_{+} \xi_{+} \xi_{+}}^{+} - 2 \beta g_{2 \xi_{-} \xi_{-} \xi_{-} \tau}^{-} + 2 \beta g_{2 \xi_{+} \xi_{+} \xi_{+} \tau}^{+} - 2 \beta f_{2 \xi_{-} \xi_{-} \xi_{-} T}^{-} + 2 \beta f_{2 \xi_{+} \xi_{+} \xi_{+} T}^{+} \notag \\
&~~~+ \gamma \lb h_{1}^{-} - h_{2}^{-} + h_{1}^{+} - h_{2}^{+} \rb - 4 \mu_{2c},
\label{S2eqpreav}
\end{align}
where again $\mu_{2c}$ is the coupling terms at this order of the expansion. Following the steps from previous orders, we average \eqref{S1eqpreav} and \eqref{S2eqpreav} at constant $\xi_{-}$ or constant $\xi_{+}$ and rearrange to obtain
\begin{equation}
	\pm 2 \psi_{1 \xi_{\pm} \tau} = H_{1}^{\pm} \lb \xi_{\pm}, \tau, T \rb + \hat{H}_{1}^{\pm} \lb \xi_{\pm}, T \rb,
	\label{psi1eqdiff}
\end{equation}
and
\begin{equation}
	\pm 2 \psi_{2 \xi_{\pm} \tau} = H_{2}^{\pm} \lb \xi_{\pm}, \tau, T \rb + \hat{H}_{2}^{\pm} \lb \xi_{\pm}, T \rb,
	\label{psi2eqdiff}
\end{equation}
where the functions $H_{1,2}^{\pm}, \hat{H}_{1,2}^{\pm}$ can be found from \eqref{S1eqpreav} and \eqref{S2eqpreav}. The equation for $\phi_{1,2}^{\pm}$ is captured by the function $\hat{H}_{1,2}^{\pm}$ and this must be zero to avoid secular terms in the same way as at previous orders. Therefore we look for terms in \eqref{S1eqpreav}, \eqref{S2eqpreav}, that depend only on $\xi_{\pm}$ and $T$. Following this approach we obtain the equations
\begin{align}
	&\lb \mp 2 \phi_{1 T}^{\pm} + \lb f_{1}^{\pm} \phi_{1}^{\pm} \rb_{\xi_{\pm}} + d_{1} \phi_{1 \xi_{\pm}}^{\pm} + \phi_{1 \xi_{\pm} \xi_{\pm} \xi_{\pm}}^{\pm} \rb_{\xi_{\pm}} = \delta \lb \phi_{1}^{\pm} - \phi_{2}^{\pm} \rb + f_{1 T T}^{\pm} \mp 2 f_{1 \xi_{\pm} \xi_{\pm} \xi_{\pm} T}^{\pm} \notag \\
	&~+ \frac{\tilde{\omega}^2 \tilde{\theta}_{1}^{2}}{2} f_{1 \xi_{\pm} \xi_{\pm}}^{\pm} -  \frac{\tilde{\theta}_{1}^{2}}{2} \lb \delta + \alpha \gamma \rb f_{2 \xi_{\pm} \xi_{\pm}}^{\pm} - \frac{\tilde{\theta}_{1}^{2}}{2} \lb f_{1 \xi_{\pm}}^{\pm^{2}} \rb_{\xi_{\pm} \xi_{\pm}},
	\label{phi1eq}
\end{align}
and
\begin{align}
	&\lb \mp 2 \phi_{2 T}^{\pm} + \alpha \lb f_{2}^{\pm} \phi_{2}^{\pm} \rb_{\xi_{\pm}} + \alpha d_{1} \phi_{2 \xi_{\pm}}^{\pm} + \frac{c^2 - 1}{\epsilon} \phi_{2 \xi_{\pm}}^{\pm} + \beta \phi_{2 \xi_{\pm} \xi_{\pm} \xi_{\pm}}^{\pm} \rb_{\xi_{\pm}} = \gamma \lb \phi_{2}^{\pm} - \phi_{1}^{\pm} \rb \notag \\
	&~+ f_{2 T T}^{\pm} \mp 2 \beta f_{2 \xi_{\pm} \xi_{\pm} \xi_{\pm} T}^{\pm} + \frac{\tilde{\omega}^2 \tilde{\theta}_{2}^{2}}{2} f_{2 \xi_{\pm} \xi_{\pm}}^{\pm} + \frac{\tilde{\theta}_{2}^{2}}{2 \alpha} \lb \delta + \alpha \gamma \rb f_{1 \xi_{\pm} \xi_{\pm}}^{\pm} - \frac{\alpha \tilde{\theta}_{2}^{2}}{2} \lb f_{2 \xi_{\pm}}^{\pm^{2}} \rb_{\xi_{\pm} \xi_{\pm}},
	\label{phi2eq}
\end{align}
where we have the modified coefficient
\begin{equation}
	\tilde{\theta}_{1} = \frac{\theta_{1}}{\sin{\lb \tilde{\omega} \tau \rb}} = \frac{d_{2} \delta}{2 \tilde{\omega}}, \quad \tilde{\theta}_{2} = \frac{\theta_{2}}{\sin{\lb \tilde{\omega} \tau \rb}} = \frac{\alpha d_{2} \gamma}{2 \tilde{\omega}}.
	\label{theta12tilde}
\end{equation}
We have now defined all functions up to and including $\O{\epsilon}$ and so stop our derivation, however the procedure could be continued to any order.

% Results
\section{Validity of the Weakly-Nonlinear Solution}
\label{sec:Validity}
In this section we perform a careful error analysis to test the validity of the constructed solution by numerically solving the equation system \eqref{ueq} - \eqref{weq} and comparing to the constructed solution \eqref{uWNL} and \eqref{wWNL} with an increasing number of terms included. To obtain this constructed solution we also need to numerically solve \eqref{f1eq}, \eqref{f2eq} for the leading order solution and \eqref{phi1eq}, \eqref{phi2eq} for the solution with terms up to and including $\O{\epsilon}$. Therefore, we use two types of numerical methods: one for the coupled Boussinesq equations and another for the coupled Ostrovsky equations (see Appendix A).

% Error analysis
%\subsection{Comparison to Constructed Solution}
We compare the solution of the coupled Boussinesq equations \eqref{ueq} - \eqref{weq} (which we shall refer to as the ``exact solution'') to the constructed weakly-nonlinear solution \eqref{uWNL} and \eqref{wWNL} with an increasing number of terms included. The parameters used for the calculations in this section are $\alpha = \beta = c = 1 + \epsilon/2$ and $\delta = \gamma = 1$. We calculate the solution in the domain $x \in [-40, 40]$ and for $t \in [0, T]$ where $T = 1/\epsilon$. The initial conditions are taken to be 
\begin{align}
F_{1}(x) &= A_{1} \sechn{2}{\frac{x}{\Lambda_{1}}} + d, \quad F_{2}(x) = A_{2} \sechn{2}{\frac{x}{\Lambda_{2}}}, \notag \\
V_{1}(x) &= 2 \frac{A_{1}}{\Lambda{1}} \sechn{2}{\frac{x}{\Lambda_{1}}},\quad V_{2}(x) = 2 c \frac{A_{2}}{\Lambda_{2}} \sechn{2}{\frac{x}{\Lambda_{2}}},
\label{uwIC}
\end{align}
where $d$ is a constant and we have $A_{1} = 6k_{1}^2$, $\Lambda_{1} = \sqrt{2}/k_{1}$, $k_{1} = 1/\sqrt{6}$, $A_{2} = 6ck_{2}^2/\alpha$, $\Lambda_{2} = \sqrt{2c\beta}/k_{2}$, $k_{2} = \sqrt{\alpha/6c}$. Here we have only added a pedestal to the initial condition for $u$ in the view of the translation symmetry of the system. In all cases considered here we have $\alpha = c$ and therefore $k_{1} = k_{2} = k = 1/\sqrt{6}$.

The comparison between the exact and weakly-nonlinear solutions at various orders of $\epsilon$ is shown in Figure \ref{fig:ErrComp1} and Figure \ref{fig:ErrComp2}. 
\begin{figure}[!htbp]
\begin{center}
	\subfigure[$\gamma = 0.1$, $d = 7$ for $u$.]{\includegraphics[width=0.4\textwidth]{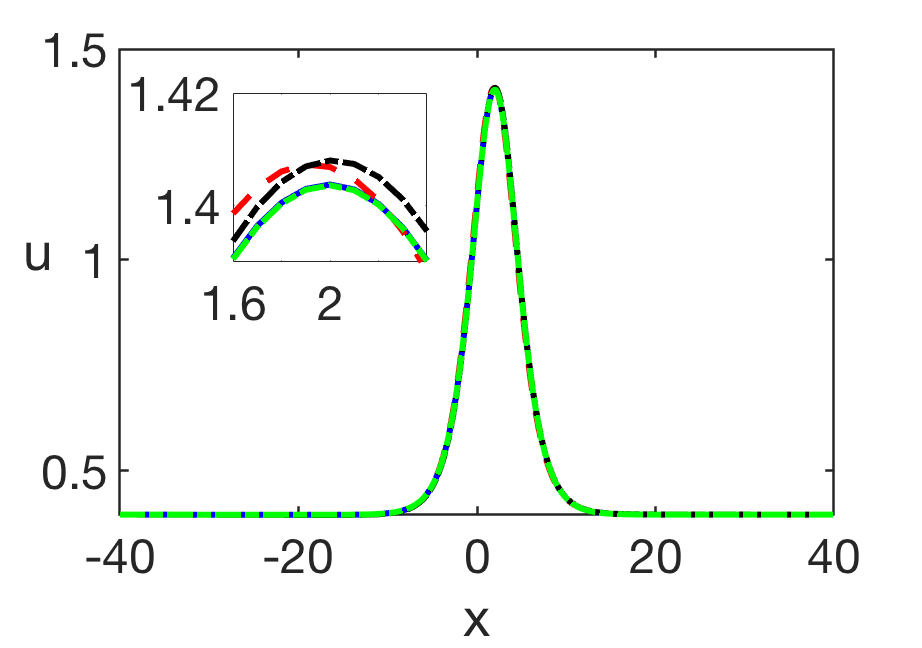}}~
	\subfigure[$\gamma = 0.1$, $d = 7$ for $w$.]{\includegraphics[width=0.4\textwidth]{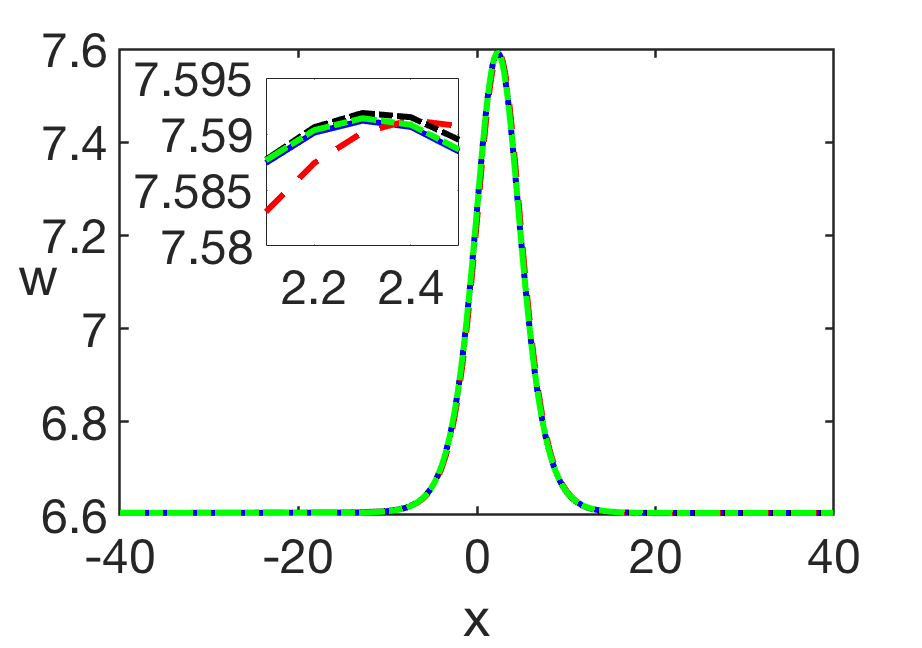}}
	\caption{\small A comparison of the direct numerical simulation (solid, blue) and the weakly-nonlinear solution including leading order (dashed, red), $\O{\sqrt{\epsilon}}$ (dash-dot, black) and $\O{\epsilon}$ (dotted, green) corrections, at $t=1/\epsilon$, for (a) $u$ and (b) $w$. Parameters are $L=40$, $N=800$, $k = 1/\sqrt{6}$, $\alpha = \beta = c = 1 + \epsilon/2$, $\gamma = 0.1$, $\epsilon = 0.0025$, $\Delta t = 0.01$ and $\Delta T = \epsilon \Delta t$. The solution agrees well to leading order, and this agreement is improved with the addition of higher-order corrections.}
	\label{fig:ErrComp1}
	\vspace{2em}
	\subfigure[$\gamma = 0.5$, $d = 7$ for $u$.]{\includegraphics[width=0.47\textwidth]{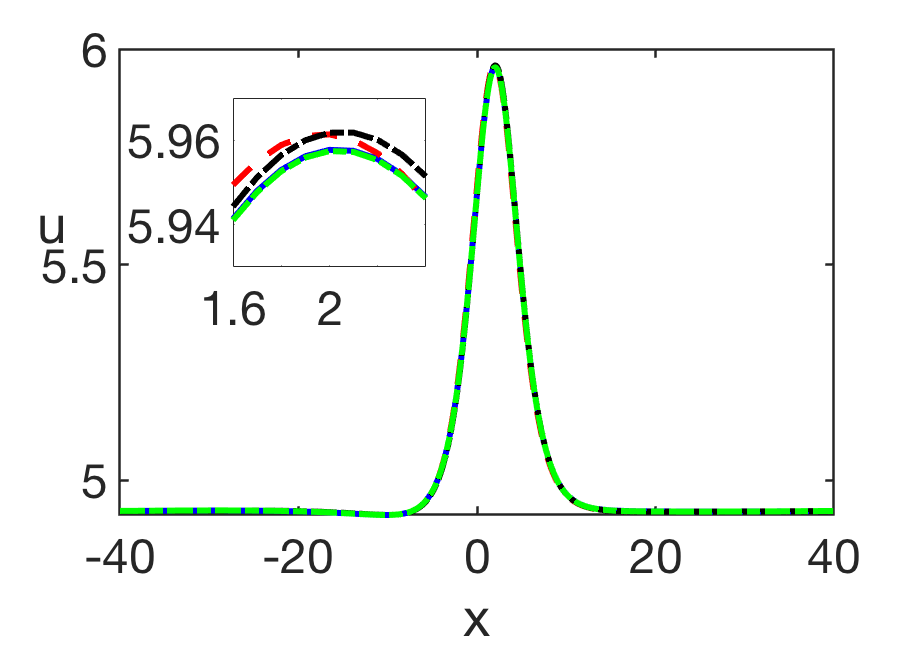}}~
	\subfigure[$\gamma = 0.5$, $d = 7$ for $w$.]{\includegraphics[width=0.47\textwidth]{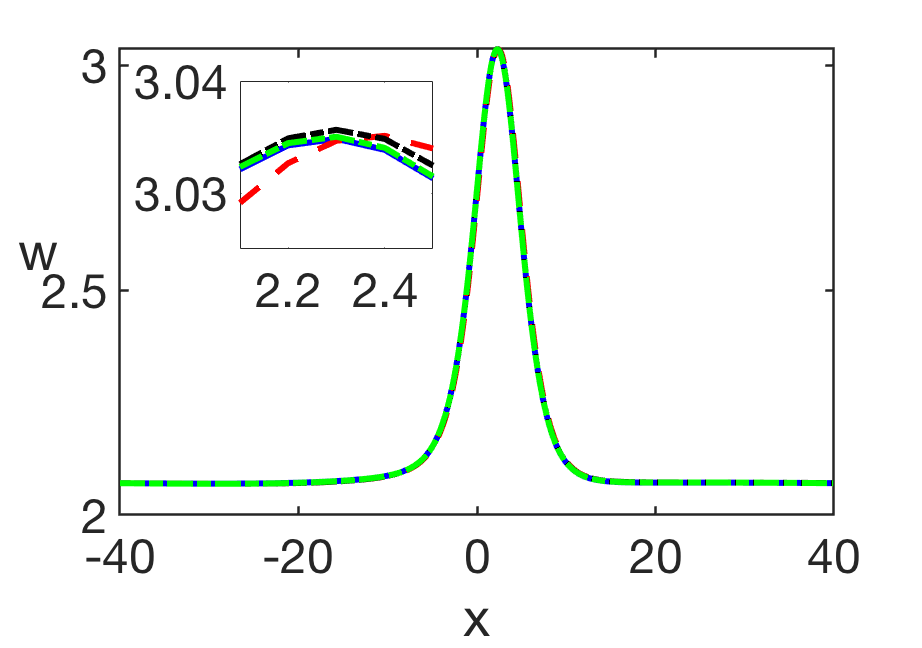}}
	\caption{\small A comparison of the direct numerical simulation (solid, blue) and the weakly-nonlinear solution including leading order (dashed, red), $\O{\sqrt{\epsilon}}$ (dash-dot, black) and $\O{\epsilon}$ (dotted, green) corrections, at $t=1/\epsilon$, for (a) $u$ and (b) $w$. Parameters are $L=40$, $N=800$, $k = 1/\sqrt{6}$, $\alpha = \beta = c = 1 + \epsilon/2$, $\gamma = 0.5$, $\epsilon = 0.0025$, $\Delta t = 0.01$ and $\Delta T = \epsilon \Delta t$. The solution agrees well to leading order and this agreement is improved with the addition of higher-order corrections.}
	\label{fig:ErrComp2}
	\end{center}
\end{figure}
We can see from the enhanced image that the leading order solution (red, dashed line) is improved with the addition of the $\O{\sqrt{\epsilon}}$ terms (black, dash-dotted line), correcting for a phase shift. The inclusion of $\O{\epsilon}$ (green, dotted line) terms adjusts the amplitude and we can see that this lies directly on top of the exact solution (blue, solid line). This is consistent for both values of $\gamma$ and $\delta$, and for both equations. We note that the larger value of $\gamma$ and $\delta$ can show a slightly increased error, however this is not as clear as the previous case for the Boussinesq equation with the Ostrovsky term in \cite{KT18}.

To understand the behaviour of the errors we denote the direct numerical solution to the system \eqref{ueq} - \eqref{weq} as $u_{\text{num}}$, the weakly-nonlinear solution \eqref{uWNL}, \eqref{wWNL}, with only the leading order terms included as $u_{1}$, with terms up to and including $\O{\sqrt{\epsilon}}$ terms as $u_{2}$ and with terms up to and including $\O{\epsilon}$ as $u_{3}$. We consider the maximum absolute error over $x$, defined as
\begin{equation}
e_{i} = \max_{-L \leq x \leq L} \abs{u_{\text{num}} \lb x, t \rb - u_{i} \lb x, t \rb}, \quad i = 1, 2, 3.
\label{MaxErr}
\end{equation}
This error is calculated at every time step and, to smooth the oscillations in the errors we average the $e_{i}$ values in the final third of the calculation, denoting this value as $\hat{e}_{i}$. We then use a least-squares power fit to determine how the maximum absolute error varies with the small parameter $\epsilon$. Therefore we write the errors in the form
\begin{equation}
\mathrm{exp} \lsq \hat{e}_{i} \rsq = C_{i} \epsilon^{\alpha_{i}},
\label{Err}
\end{equation}
and take the logarithm of both sides to form the error plot (the exponential factor is included so that we have $\hat{e}_{i}$ as the plotting variable). The values of $C_{i}$ and $\alpha_{i}$ are found using the MATLAB function \textit{polyfit}.

The corresponding errors for the cases considered in Figure \ref{fig:ErrComp1} and Figure \ref{fig:ErrComp2}  are plotted in Figure \ref{fig:ErrPlots1} and Figure \ref{fig:ErrPlots2}, that is for $\gamma = 0.1$, $\gamma = 0.5$ and $d = 7$. We can see that there is an excellent correlation for each of the curves and that the errors improve with the addition of more terms in the expansion. In the case of $\gamma = 0.1$ the slope of the error curves is 0.52, 1.00, 1.52 for $u$ and 0.50, 1.00 and 1.50 for $w$, for the inclusion of leading order, $\O{\sqrt{\epsilon}}$ and $\O{\epsilon}$ terms in the expansion respectively, in close agreement with the theoretical values. When $\gamma = 0.5$ the slope of the curves is approximately 0.50, 1.00 and 1.63 for $u$ and 0.53, 1.00 and 1.68 for $w$, showing that as $\gamma$ increases the slope of the error curve for the approximation including $\O{\epsilon}$ terms increases.
\begin{figure}[!htbp]
\begin{center}
	\subfigure[$\gamma = 0.1$, $d = 7$ for $u$.]{\includegraphics[width=0.47\textwidth]{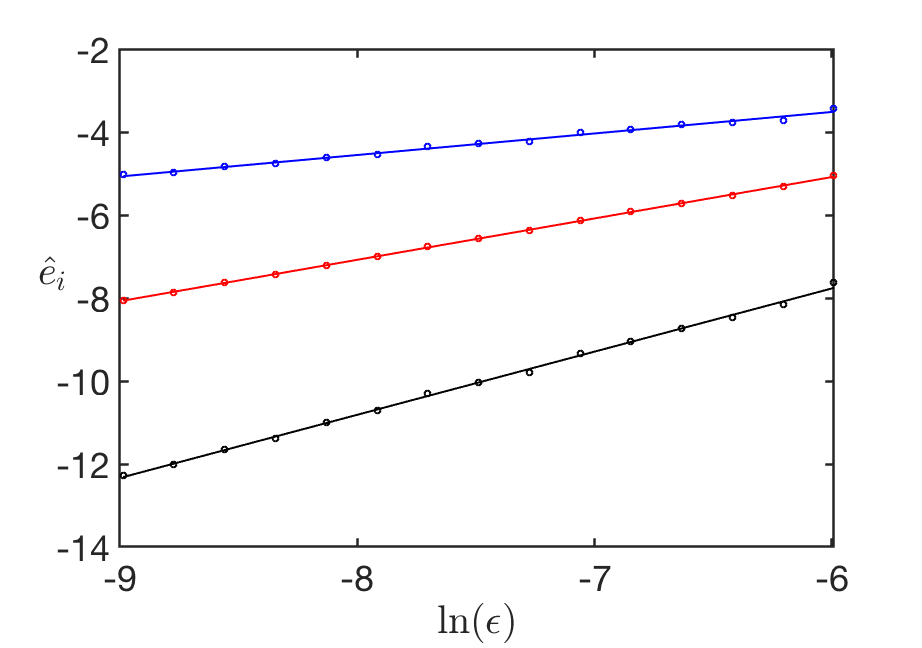}}~
	\subfigure[$\gamma = 0.1$, $d = 7$ for $w$.]{\includegraphics[width=0.47\textwidth]{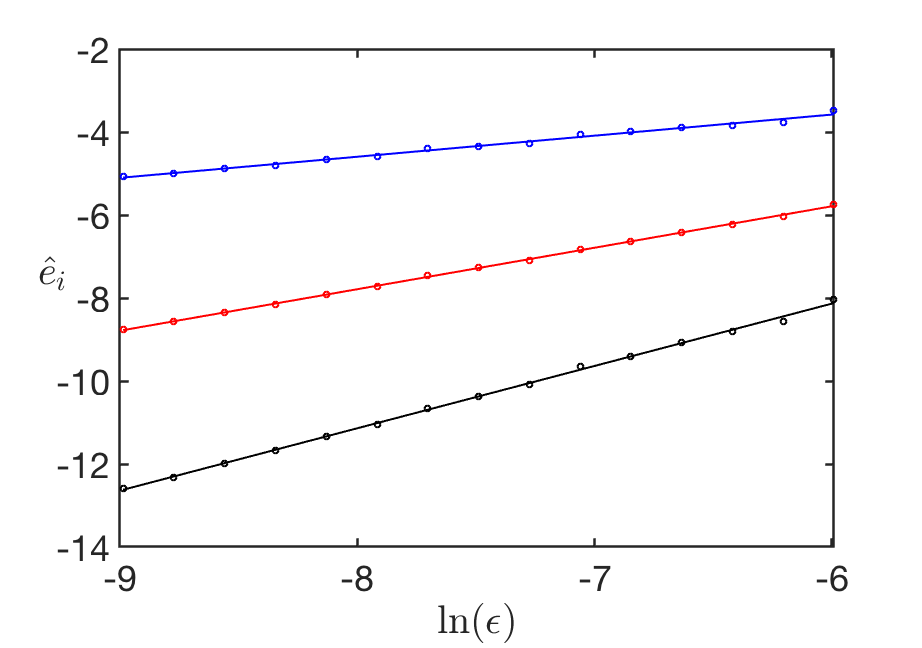}}
	\caption{\small A comparison of error curves for varying values of $\epsilon$, at $t=1/\epsilon$, for the weakly-nonlinear solution including leading order (upper, blue), $\O{\sqrt{\epsilon}}$ (middle, red) and $\O{\epsilon}$ (lower, black) corrections, for (a) $u$ and (b) $w$. Parameters are $L=40$, $N=800$, $k = 1/\sqrt{6}$, $\alpha = \beta = c = 1 + \epsilon/2$, $\gamma = 0.1$, $\Delta t = 0.01$ and $\Delta T = \epsilon \Delta t$. The inclusion of more terms in the expansion increases the accuracy.}
	\label{fig:ErrPlots1}
	\vspace{2em}
	\subfigure[$\gamma = 0.5$, $d = 7$ for $u$.]{\includegraphics[width=0.47\textwidth]{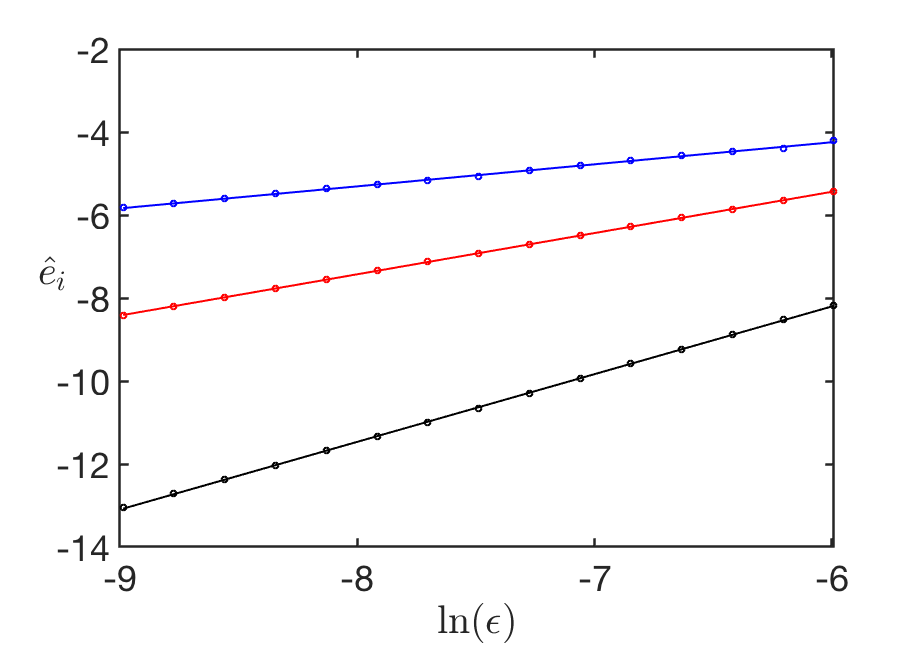}}~
	\subfigure[$\gamma = 0.5$, $d = 7$ for $w$.]{\includegraphics[width=0.47\textwidth]{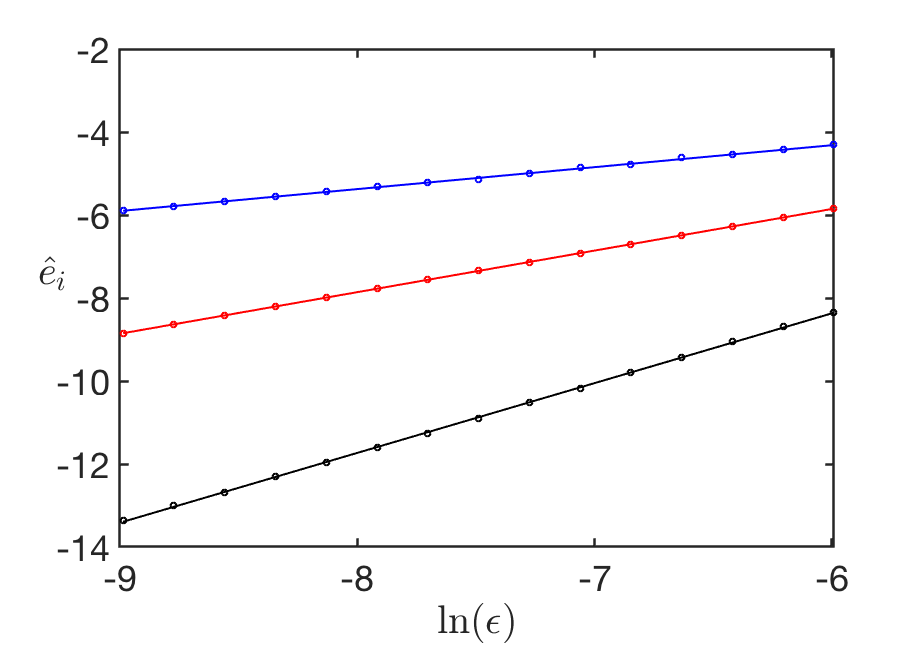}}
	\caption{\small A comparison of error curves for varying values of $\epsilon$, at $t=1/\epsilon$, for the weakly-nonlinear solution including leading order (upper, blue), $\O{\sqrt{\epsilon}}$ (middle, red) and $\O{\epsilon}$ (lower, black) corrections, for (a) $u$ and (b) $w$. Parameters are $L=40$, $N=800$, $k = 1/\sqrt{6}$, $\alpha = \beta = c = 1 + \epsilon/2$, $\gamma = 0.5$, $\Delta t = 0.01$ and $\Delta T = \epsilon \Delta t$. The inclusion of more terms in the expansion increases the accuracy.}
	\label{fig:ErrPlots2}
	\end{center}
\end{figure}

% Radiating solitary wave
\section{Radiating Solitary Waves}
\label{sec:RSW}
In this section we study the interaction of two radiating solitary waves, using both the direct numerical simulations and the weakly-nonlinear solution. Firstly we consider the case of a single radiating solitary wave and show that we have a good agreement between the direct numerical simulation and the weakly-nonlinear solution, then we use both methods to reliably study the complicated case of two interacting radiating solitary waves.

In order to better resolve the tail of a radiating solitary wave,  we scale the variables as $u = 12 \tilde{u}$, $w = 12 \tilde{w}$, so that we obtain (omitting tildes)
\begin{align}
&u_{tt} - u_{xx} = \epsilon \lsq 6 \lb u^2 \rb_{xx} + u_{ttxx} - \delta \lb u - w \rb \rsq, \label{ueqscaled} \\
&w_{tt} - c^2 w_{xx} = \epsilon \lsq 6 \alpha \lb w^2 \rb_{xx} + \beta w_{ttxx} + \gamma \lb u - w \rb \rsq. \label{weqscaled}
\end{align}
The weakly-nonlinear solution can be easily scaled as well so we obtain, for $f_{1,2}^{\pm}$,
\begin{align}
	&\lb \mp 2 f_{1 T}^{\pm} + 12 f_{1}^{\pm} f_{1 \xi_{\pm}}^{\pm} + d_{1} f_{1 \xi_{\pm}}^{\pm} + f_{1 \xi_{\pm} \xi_{\pm} \xi_{\pm}}^{\pm} \rb_{\xi_{\pm}} = \delta \lb f_{1}^{\pm} - f_{2}^{\pm} \rb, \notag \\
	&\lb \mp 2 f_{2 T}^{\pm} + 12 \alpha f_{2}^{\pm} f_{2 \xi_{\pm}}^{\pm} + \lb \alpha d_{1} + \frac{c^2 - 1}{\epsilon} \rb f_{2 \xi_{\pm}}^{\pm} + \beta f_{2 \xi_{\pm} \xi_{\pm} \xi_{\pm}}^{\pm} \rb_{\xi_{\pm}} = \gamma \lb f_{2}^{\pm} - f_{1}^{\pm} \rb.
	\label{fcOstScaled}
\end{align}
Similarly for the function $g_{1,2}^{\pm}$ we have
\begin{equation}
	g_{1}^{\pm} = \pm \frac{6 d_{2} \delta}{\tilde{\omega}} \sin{\lb \tilde{\omega} \tau \rb} f_{1 \xi_{\pm}}^{\pm},
	\label{g1eqScaled}
\end{equation}
and
\begin{equation}
	g_{2}^{\pm} = \mp \frac{6 \alpha d_{2} \gamma}{\tilde{\omega}} \sin{\lb \tilde{\omega} \tau \rb} f_{2 \xi_{\pm}}^{\pm}.
	\label{g2eqScaled}
\end{equation}
In this section we only use the weakly-nonlinear solution up to and including $\O{\sqrt{\epsilon}}$ terms so we do not consider the effect of the scaling on the higher-order terms. We will show that even this approximation produces good qualitative and reasonable quantitative results, even in very long runs. 

To show the agreement between the direct numerical simulations and the weakly-nonlinear solution for a radiating solitary wave, we take the initial condition to be the solitary wave solution of the uncoupled Boussinesq equation for $u$ and $w$, as presented in \eqref{uwIC}. Therefore we have $f_{1,2}^{+} = 0$. The parameters in the equation are $\alpha = \beta = c = 1 + \epsilon/2$, $\gamma = 1$ and $\epsilon = 0.01$. The domain size is taken to be $L = 1000$ and therefore we have $N = 20000$, and the time step is $\Delta t = 0.01$ as before. We take the phase shift in the initial conditions to be $x_{0} = 800$ and the result of the computation at $t = 1400$ is presented in Figure \ref{fig:RSW}. We can see that the radiating solitary wave is formed in both equations and that the agreement between the direct numerical simulations (blue, solid line) and weakly-nonlinear solution (red, dashed line) is good, with a small phase shift between the two solutions and a small discrepancy in the amplitude.
\begin{figure}[!htbp]
	\begin{center}
		\subfigure[Solution for $u$.]{\includegraphics[width=0.47\textwidth]{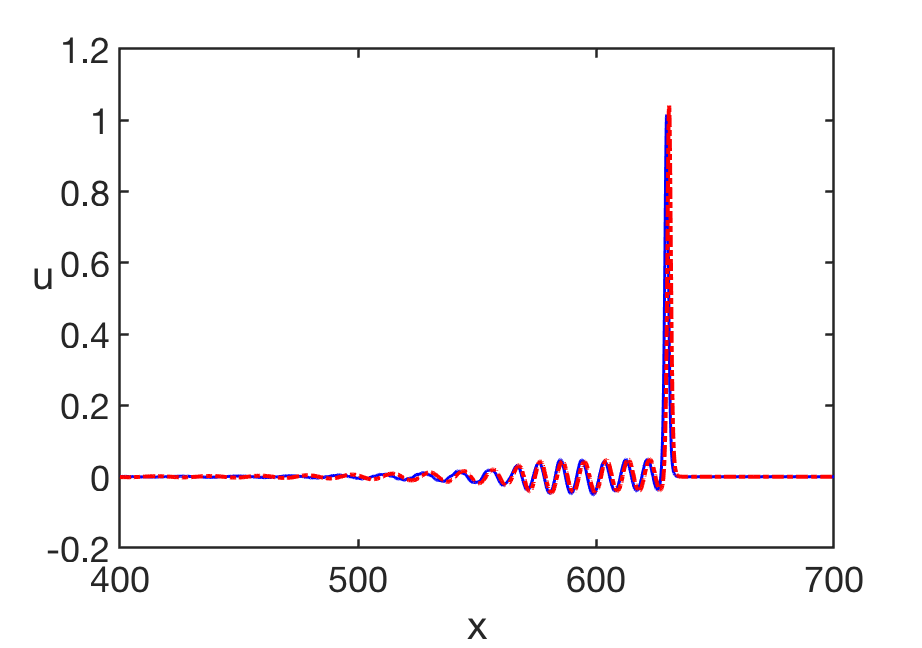}}~
		\subfigure[Solution for $w$.]{\includegraphics[width=0.47\textwidth]{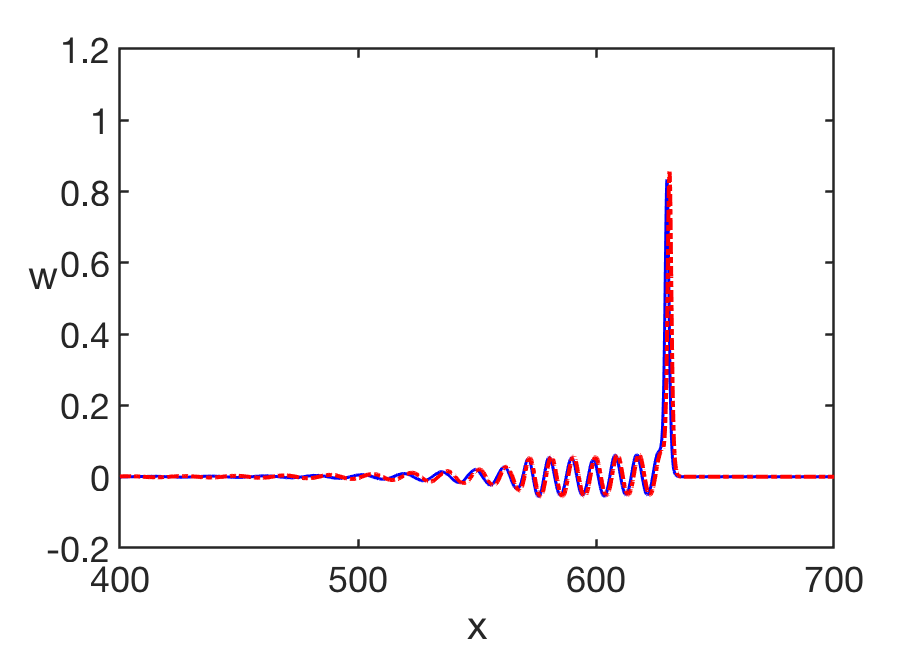}}
		\caption{\small A comparison of the numerical solution (solid, blue) at $t=1400$ and the weakly-nonlinear solution including $\O{\sqrt{\epsilon}}$ terms (dash-dot, red), for (a) $u$ and (b) $w$. Parameters are $L=1000$, $N=20000$, $k = 1/\sqrt{6}$, $\alpha = \beta = c = 1 + \epsilon/2$, $\gamma = 1$, $\epsilon = 0.01$, $\Delta t = 0.01$ and $\Delta T = \epsilon \Delta t$. 
		%The solution agrees reasonably well to leading order, and this agreement is improved with the addition of higher order corrections.
		}
		\label{fig:RSW}
	\end{center}
		\end{figure}

We now consider the case when two radiating solitary waves interact. To obtain an appropriate initial condition we use the two-soliton solution for the KdV equation as the initial condition for the coupled Boussinesq equations and choose the second initial condition in the appropriate form so that there is no left-propagating wave, as was done in \cite{KM12}. Explicitly we take
\begin{align}
	u(x,0) &= \frac{\lb k_{1} - k_{2} \rb^2 + \sqrt{C} \lb k_{1}^2 \cosh{\lb k_{2}x + x_{1} \rb} + k_{2}^{2} \cosh{\lb k_{1}x + x_{0} \rb} \rb}{2 \lsq \cosh{\lb \lb k_{1}x - k_{2} x + x_{0} - x_{1} \rb/2 \rb} + \sqrt{C} \cosh{\lb \lb k_{1}x + k_{2} x + x_{0} + x_{1} \rb/2 \rb} \rsq^2}, \notag \\
	u_{t}(x,0) &= -\diff{ }{x} u(x,0),
	\label{KdV2SolIC}
\end{align}
where $C = \lsq \lb k_{1} - k_{2} \rb/\lb k_{1} + k_{2} \rb \rsq^2$ and we take the same initial condition for $w$. In what follows, to ensure the radiating solitary waves have enough time to interact, we use a large domain and a long time for the calculation. Therefore we take $L = 5000$ and $N = 100000$ for the calculation, and the parameters are $\alpha = \beta = c = 1 + \epsilon/2$, $\gamma = 1$, $\epsilon = 0.01$. To ensure that the initial solitons are well separated and have sufficiently different amplitudes, we take $k_{1} = 1$, $k_{2} = 2$, $x_{0} = -50$ and $x_{1} = 50$. The results of the calculation at various times are presented in Figure \ref{fig:RSWInt}. 
\begin{figure}[!htbp]
	\begin{center}
%		\subfigure[Solution at $t = 0$.]{\includegraphics[width=0.47\textwidth]{RSWIntt0}}~
		\subfigure[Solution for $u$ at $t = 800$.]{\includegraphics[width=0.445\linewidth]{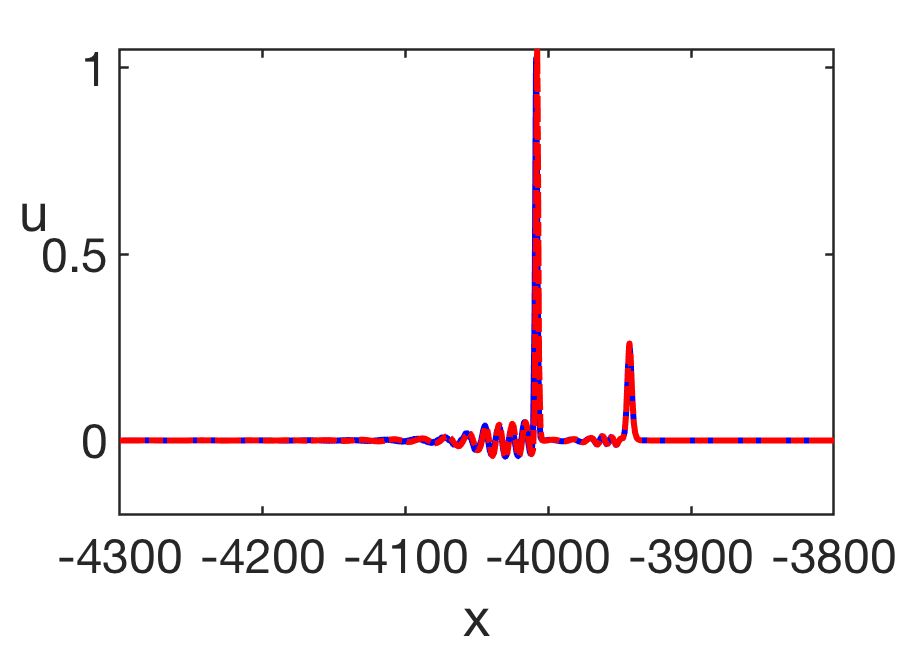}}
		\subfigure[Solution for $w$ at $t = 800$.]{\includegraphics[width=0.445\linewidth]{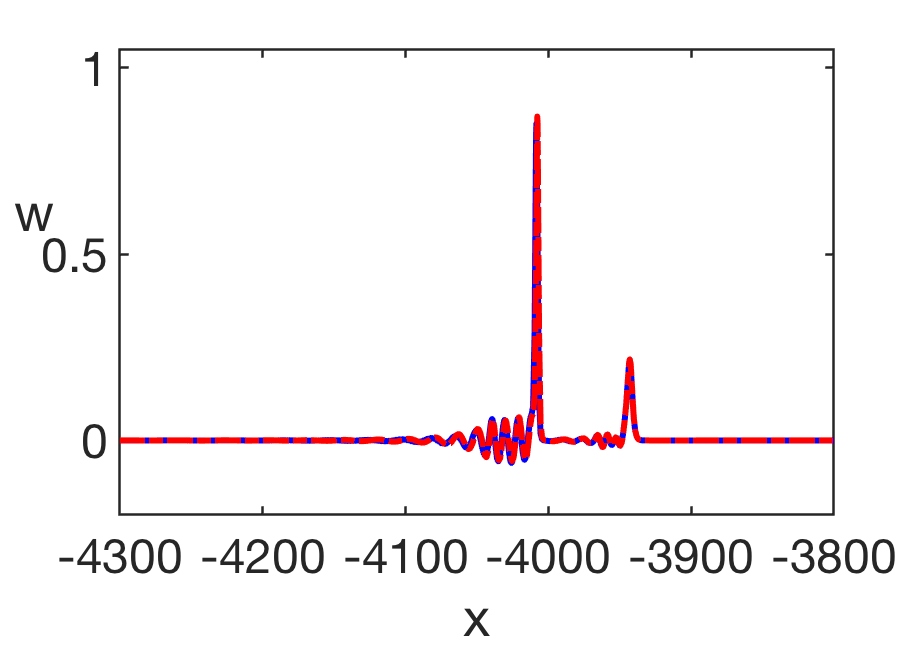}}	
		\subfigure[Solution for $u$ at $t = 5800$.]{\includegraphics[width=0.445\linewidth]{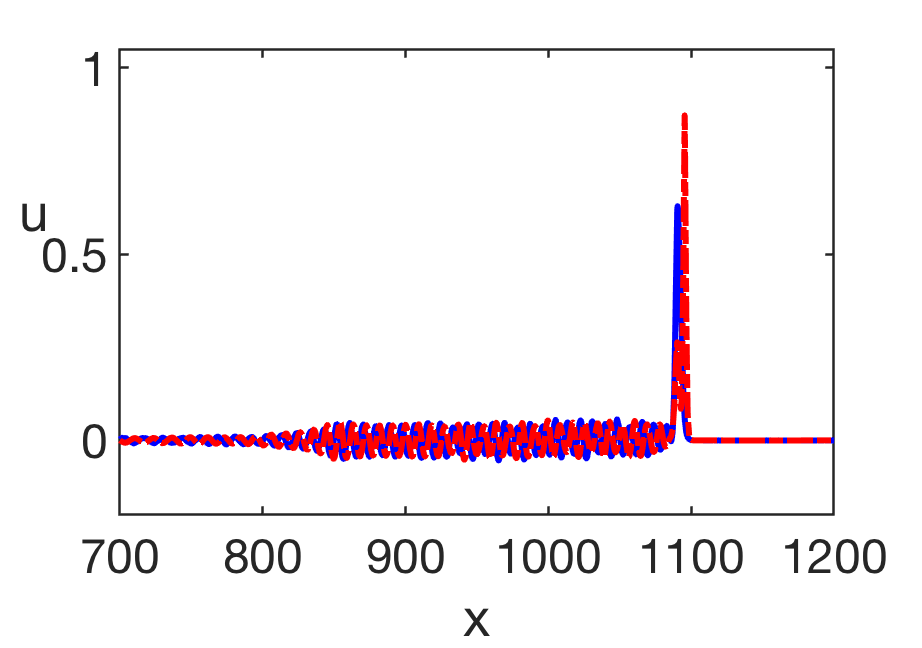}} 
		\subfigure[Solution for $w$ at $t = 5800$.]{\includegraphics[width=0.445\linewidth]{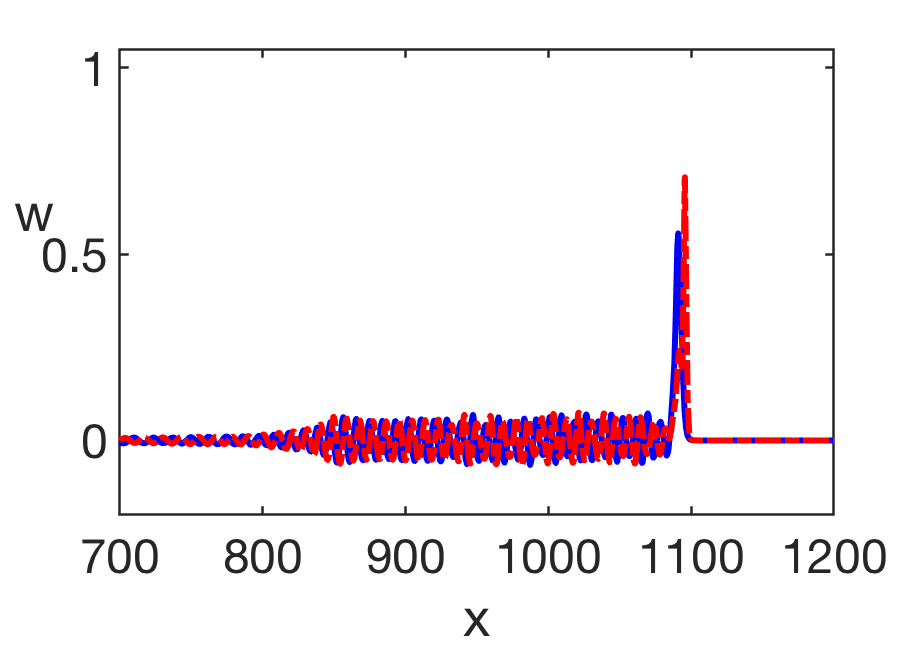}}	
		\subfigure[Solution for $u$ at $t = 9500$.]{\includegraphics[width=0.445\linewidth]{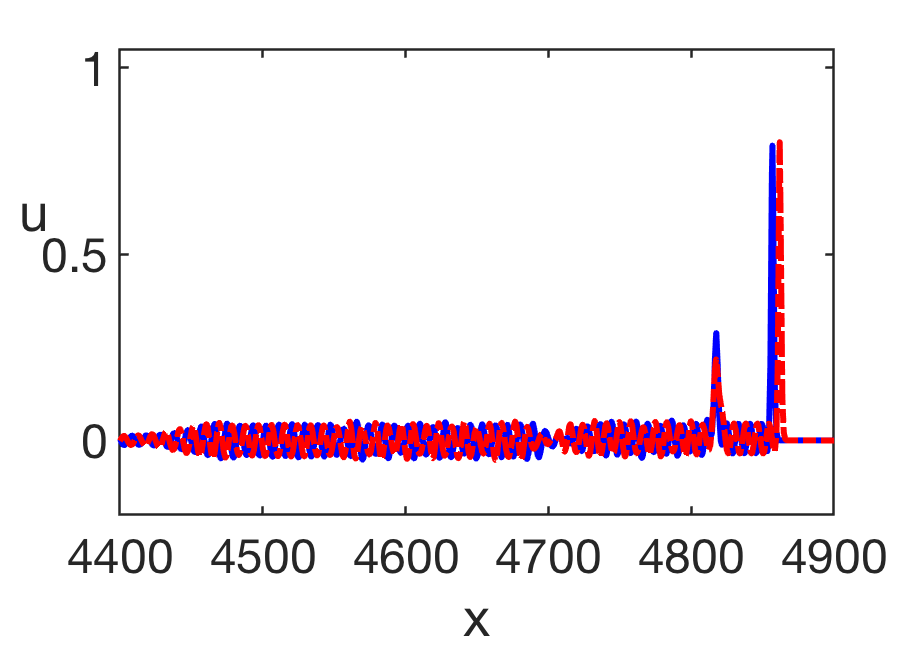}} 
		\subfigure[Solution for $w$ at $t = 9500$.]{\includegraphics[width=0.445\linewidth]{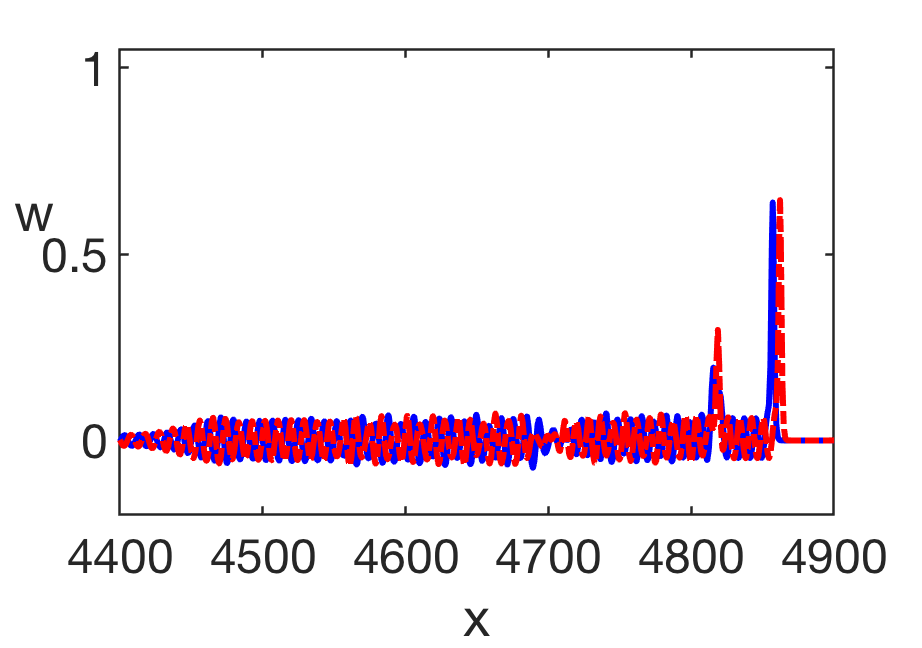}}	
		\caption{\small A comparison of the direct numerical solution  (solid, blue) and the weakly-nonlinear solution including $\O{\sqrt{\epsilon}}$ terms  (dashed, red) at various times for (a), (c), (e) $u$  and (b), (d), (f) $w$. Parameters are $L = 5000$, $N = 100000$, $k_{1} = 1$, $k_{2} = 2$, $x_{0} = -50$, $x_{1} = 50$, $\alpha = \beta = c = 1 + \epsilon/2$, $\gamma = 1$, $\epsilon = 0.01$, $\Delta t = 0.01$ and $\Delta T = \epsilon \Delta t$. The difference between the amplitudes of the non-stationary solitons during and after the interaction is cased by a small phase shift between the exact and weakly-nonlinear solutions.
	%The interaction of the radiating solitary waves results in a decrease in amplitude and the generation of a wave packet from the radiating tails before the interaction.
		}
		\label{fig:RSWInt}
	\end{center}
\end{figure}

We can see that the weakly-nonlinear solution (red, dashed line) agrees very well with the results of direct numerical simulations (blue, solid line), even in such complicated interaction problem, and the two approaches verify each other. There is only a small phase shift, which becomes more noticeable at larger times.  At $t = 800$ the larger soliton has begun to travel through the tail of the smaller soliton and the amplitude has begun to reduce as it travels through the tail. At $t = 5800$, the two solitons now lie on top of each other (in direct numerical simulations; the weakly-nonlinear solution has a small phase shift). We can see that the amplitude is reduced in contrast to the initial condition, and the tail now shows the presence of two frequencies. At $t = 9500$, the larger radiating solitary wave has overtaken the smaller wave,  and we again have two distinct radiating solitary waves with tails, although with a significantly reduced amplitude. Furthermore, we can see the formation of a wave packet behind the smaller soliton which was generated by the interaction of the solitons. The disagreement between the amplitudes of the smaller soliton after the interaction is again a consequence of a small phase shift, since the amplitude of the smaller soliton keeps oscillating between $u$ and $w$ as wave tries to settle after the interaction. This has been verified by comparing the maximum of the amplitude of the smaller soliton in the two approaches in a small interval around the fixed moment of time (over the period of oscillations). The difference is of order $\O{\epsilon}$, in agreement with the accuracy of the approximation used in this section.

\section{Conclusions}
\label{sec:Conc}

In this paper we generalised the results of our recent study \cite{KT18}, where we developed a new asymptotic procedure for the construction of the d'Alembert-type solution of the Cauchy problem for a Boussinesq-type equation with the Ostrovsky term. We have shown that the developed approach can be extended to construct a similar solution for a system of coupled Boussinesq equations (\ref{ueq}), (\ref{weq}), describing long longitudinal strain waves in a bi-layer with an imperfect interface \cite{KSZ}. We examined the accuracy of the constructed solution numerically, and we used both the direct numerical simulations for the coupled Boussinesq equations, and our constructed  semi-analytical solution in order to study the complicated process of the interaction of two radiating solitary waves. The two approaches showed excellent agreement, even in very long runs. The constructed solution can find useful applications in the studies of the scattering of radiating solitary waves by delamination \cite{KT17} and other extended inhomogeneities.

\section{Acknowledgements}
\label{sec:Ackn}

We thank D.E. Pelinovsky and A.V. Porubov for useful discussions and references. KRK is grateful to the UK QJMAM Fund for Applied Mathematics for the support of her travel to the ESMC2018 in Bologna, Italy where some of these discussions have taken place. MRT is grateful to the UK Institute of Mathematics and its Applications and the London Mathematical Society for supporting travel to the same conference.

\titleformat{\section}{\large\bfseries}{\appendixname~\thesection :}{0.5em}{}
\appendix
\section{Numerical Methods}
\label{sec:NumMeth}

% Numerical methods used here for cRB and cOst
%\subsection{Numerical Methods}
In the following methods we use the Discrete Fourier Transform (DFT) to calculate the Fourier transform of numerical data (e.g., \cite{T}). Let us consider a function $u(x,t)$ on a finite domain $x \in [-L, L]$ and we discretise the domain into $N$ equally spaced points, so we have the spacing $\Delta x = 2L/N$. In all calculations  we scale the domain from $x \in [-L, L]$ to $\tilde{x} \in [0, 2\pi]$, which can be achieved by applying the transform $\tilde{x} = sx + \pi$, where $s = \pi/L$. Denoting $x_{j} = -L + j \Delta x$ for $j=0,\dots,N$, we define the DFT for the function $u(x,t)$ as
\begin{equation}
\hat{u} \lb k, t \rb = \frac{1}{\sqrt{N}} \sum_{j=1}^{N} u \lb x_{j}, t \rb e^{-i k x_{j}}, \  -\frac{N}{2} \leq k \leq \frac{N}{2} - 1,
\label{DFT}
\end{equation}
and similarly the IDFT is defined as
\begin{equation}
u \lb x, t \rb = \frac{1}{\sqrt{N}} \sum_{k=-N/2}^{N/2 - 1} \hat{u} \lb k, t \rb e^{i k x_{j}}, \quad j =  1, 2, \dots, N,
\label{IDFT}
\end{equation}
where the discretised and scaled wavenumber is now $k \in \mathbb{Z}$. To perform these transforms we implement the FFTW3 algorithm in C \cite{FFTW3}. 

For the coupled Boussinesq equations \eqref{ueq} - \eqref{weq} we use a pseudospectral method similar to the one presented in \cite{EST}, where this method was used to solve a single regularised Boussinesq equation in the context of microstructured solids. We introduce the change of variable
\begin{equation}
U = u - \epsilon u_{xx}, \quad W = w - \epsilon \beta w_{xx},
\label{PSUW}
\end{equation}
so that we have
\begin{align}
&U_{tt} = u_{xx} + \epsilon \lsq \frac{1}{2} \lb u^2 \rb_{xx} - \delta \lb u - w \rb \rsq, \notag \\
&W_{tt} = c^2 w_{xx} + \epsilon \lsq \frac{\alpha}{2} \lb w^2 \rb_{xx} + \gamma \lb u - w \rb \rsq.
\label{PScRB}
\end{align}
Taking the Fourier transform of \eqref{PSUW} we obtain
\begin{equation}
\hat{u} = \frac{\hat{U}}{1 + \epsilon k^2}, \quad \hat{w} = \frac{\hat{W}}{1 + \epsilon \beta k^2}.
\label{PSuwTransform}
\end{equation}
We take the Fourier transform of \eqref{PScRB} and substitute \eqref{PSuwTransform} into this expression to obtain an ODE in $\hat{u}$ and $\hat{w}$, taking the form
\begin{align}
\hat{U}_{tt} &= -\frac{\epsilon \delta + s^2 k^2}{1 + \epsilon s^2 k^2} \hat{U} - \frac{\epsilon s^2 k^2}{2} \mathscr{F} \lset \mathscr{F}^{-1} \lsq \frac{\hat{U}}{1 + \epsilon s^2 k^2} \rsq^2 \rset + \frac{\epsilon \delta}{1 + \epsilon s^2 \beta k^2} \hat{W}^2, = \hat{S}_1 \lb \hat{U}, \hat{W} \rb, \notag \\
\hat{W}_{tt} &= -\frac{\epsilon \gamma + c^2 s^2 k^2}{1 + \epsilon \beta s^2 k^2} \hat{W} - \frac{\epsilon \alpha s^2 k^2}{2} \mathscr{F} \lset \mathscr{F}^{-1} \lsq \frac{\hat{W}}{1 + \epsilon \beta s^2 k^2} \rsq^2 \rset + \frac{\epsilon \delta}{1 + \epsilon s^2 k^2} \hat{U}^2 = \hat{S}_{2} \lb \hat{U}, \hat{W} \rb,
\label{PSuwODE}
\end{align}
where $\mathscr{F}$ denotes the Fourier transform. We solve this system of ODEs using a 4$^{\mathrm{th}}$-order Runge-Kutta method for time stepping, such as the one used in \cite{KT17}. Let us rewrite the system as
\begin{align}
&\hat{U}_{t} = \hat{G}, &&\hat{W}_{t} = \hat{H}, \notag \\
&\hat{G}_{t} = \hat{S}_1 \lb \hat{U}, \hat{W} \rb, &&\hat{H}_{t} = \hat{S}_2 \lb \hat{U}, \hat{W} \rb,
\label{PScRBSplit}
\end{align}
where we defined $\hat{S}_{1,2}$ as the right-hand side of \eqref{PSuwODE}. We discretise the time domain and functions as $t = t_n$, $\hat{U}(k, t_{n}) = \hat{U}_n$, $\hat{W}(k, t_{n}) = \hat{W}_n$, $\hat{G}(k, t_{n}) = \hat{G}_{n}$, $\hat{H}(k, t_{n}) = \hat{H}_{n}$ for $n=0,1,2,\dots$, where $t_{n} = n \Delta t$, and $k$ discretises the Fourier space. Taking the Fourier transform of the initial conditions as defined in \eqref{uIC} and \eqref{wIC}, and making use of \eqref{PSuwTransform}, we obtain initial conditions $\hat{U}_{0}$, $\hat{W}_{0}$, and $\hat{G}_0$, $\hat{H}_{0}$ of the form
\begin{align}
&\hat{U}_0 = \lb 1 + \epsilon s^2 k^2 \rb \mathscr{F} \lset F_{1}(x) \rset, &&\hat{W}_0 = \lb 1 + \epsilon \beta s^2 k^2 \rb \mathscr{F} \lset F_{2}(x) \rset, \notag \\
&\hat{G}_0 = \lb 1 + \epsilon s^2 k^2 \rb \mathscr{F} \lset V_{1}(x) \rset, &&\hat{H}_0 = \lb 1 + \epsilon \beta s^2 k^2 \rb \mathscr{F} \lset V_{2}(x) \rset.
\label{PSUWIC}
\end{align}
Now we have initial conditions, we implement a 4$^{\mathrm{th}}$-order Runge-Kutta method, taking the form 
\begin{align*}
&\hat{U}_{n+1} = \hat{U}_n + \frac{1}{6} \lsq k_1 + 2k_2 + 2k_3 + k_4 \rsq, &&\hat{G}_{n+1} = \hat{G}_n + \frac{1}{6} \lsq l_1 + 2l_2 + 2l_3 + l_4 \rsq, \\
&\hat{W}_{n+1} = \hat{W}_n + \frac{1}{6} \lsq m_1 + 2m_2 + 2m_3 + m_4 \rsq, &&\hat{H}_{n+1} = \hat{H}_n + \frac{1}{6} \lsq n_1 + 2n_2 + 2n_3 + n_4 \rsq,
\end{align*}
where
\begin{align}
&k_{1} = \Delta t \hat{G}_n, &&l_{1} = \Delta t \hat{S}_{1} \lb \hat{U}_{n}, \hat{W}_{n} \rb, \notag \\
&m_{1} = \Delta t \hat{H}_n, &&n_{1} = \Delta t \hat{S}_{2} \lb \hat{U}_{n}, \hat{W}_{n} \rb, \notag \\
&k_{2} = \Delta t \lb \hat{G}_n + \frac{l_{1}}{2} \rb, &&l_{2} = \Delta t \hat{S}_{1} \lb \hat{U}_{n} + \frac{k_{1}}{2}, \hat{W}_{n} + \frac{m_{1}}{2} \rb, \notag \\
&m_{2} = \Delta t \lb \hat{H}_n + \frac{n_{1}}{2} \rb, &&n_{2} = \Delta t \hat{S}_{2} \lb \hat{U}_{n} + \frac{k_{1}}{2}, \hat{W}_{n} + \frac{m_{1}}{2} \rb, \notag \\
&k_{3} = \Delta t \lb \hat{G}_n + \frac{l_{2}}{2} \rb, &&l_{3} = \Delta t \hat{S}_{1} \lb \hat{U}_{n} + \frac{k_{2}}{2}, \hat{W}_{n} + \frac{m_{2}}{2} \rb, \notag \\
&m_{3} = \Delta t \lb \hat{H}_n + \frac{n_{2}}{2} \rb, &&n_{3} = \Delta t \hat{S}_{2} \lb \hat{U}_{n} + \frac{k_{2}}{2}, \hat{W}_{n} + \frac{m_{2}}{2} \rb, \notag \\
&k_{4} = \Delta t \lb \hat{G}_n + l_{3} \rb, &&l_{4} = \Delta t \hat{S}_{1} \lb \hat{U}_{n} + k_{3}, \hat{W}_{n} + m_{3} \rb, \notag \\
&m_{4} = \Delta t \lb \hat{H}_n + n_{3} \rb, &&n_{4} = \Delta t \hat{S}_{2} \lb \hat{U}_{n} + k_{3}, \hat{W}_{n} + m_{3} \rb.
\label{PSBOstRK4}
\end{align}
To obtain the solution in the real domain, we transform $\hat{U}$ back to $u$, and similarly $\hat{W}$ back to $w$, through relation \eqref{PSuwTransform}. Explicitly we have
\begin{equation}
u(x,t) = \mathscr{F}^{-1} \lset \frac{\hat{U}}{1 + \epsilon s^2 k^2} \rset, \quad w(x,t) = \mathscr{F}^{-1} \lset \frac{\hat{W}}{1 + \epsilon s^2 \beta k^2} \rset.
\label{PSUWReal}
\end{equation}

We now consider the solution to the coupled Ostrovsky equations. The method is similar to that used in \cite{AGK}. It is presented for the equations \eqref{phi1eq}, \eqref{phi2eq}, as this method can be reduced to solve the system \eqref{f1eq}, \eqref{f2eq}. We present the equations for the negative superscript i.e. for $\phi_{1}^{-}$ and $\phi_{2}^{-}$. We omit the superscript in the subsequent equations. Let us consider the system of coupled Ostrovsky equations defined as
\begin{align}
\lb \phi_{1 t} + \omega_{1} \phi_{1 x} + \alpha_1 \lb f_{1} \phi_{1} \rb_{x} + \beta_1 \phi_{1 xxx} \rb_{x} &= \delta \lb \phi_{1} - \phi_{2} \rb + H_{1} \lb f_{1}(x), f_{2}(x) \rb, \notag \\
\lb \phi_{2 t} + \omega_{2} \phi_{2 x} + \alpha_2 \lb f_{2} \phi_{2} \rb_{x} + \beta_2 \phi_{2 xxx}  \rb_{x} &= \gamma \lb \phi_{2} - \phi_{1} \rb + H_{2} \lb f_{1}(x), f_{2}(x) \rb,
\label{PScOst}
\end{align}
where $\alpha_1$, $\alpha_2$, $\beta_1$, $\beta_2$, $\omega_{1}$, $\omega_{2}$, $\delta$ and $\gamma$ are constants, and the functions $f_{1}$, $f_{2}$ are known. We consider the equation on domains $t \in [0, T]$ and $x \in [-L, L]$. We calculate the nonlinear terms in the real domain and then transform them to the Fourier space. The spatial domain is discretised by $N$ equidistant points with spacing $\Delta x = 2 \pi / N$, and we have the DFT and IDFT as defined in \eqref{DFT} and \eqref{IDFT} respectively, with an appropriately similar transform for $w$. The discrete Fourier transform of equations \eqref{PScOst} with respect to $x$ gives
\begin{align}
\hat{\phi}_{1t} + \lb i s k \omega_{1} - i s^3 k^3 \beta_1 \rb \hat{\phi}_{1} + i s k \alpha_1 \mathscr{F} \lset f_{1} \phi_{1} \rset  &= -\frac{i \delta}{sk} \lb \hat{\phi}_{1} - \hat{\phi}_{2} \rb - \frac{i}{sk} \hat{H}_{1}, \notag \\
\hat{\phi}_{2t} + \lb i s k \omega_{2} - i s^3 k^3 \beta_{2} \rb \hat{\phi}_{2} + i s k \alpha_2 \mathscr{F} \lset f_{2} \phi_{2} \rset &= -\frac{i \gamma}{sk} \lb \hat{\phi}_{2} - \hat{\phi}_{1} \rb - \frac{i}{sk} \hat{H}_{2}.
\label{PScOstFFT}
\end{align}
This system is solved numerically using a 4$^\mathrm{th}$-order Runge-Kutta method for time stepping as for the coupled Boussinesq equations. Assume that the solution at $t$ is given by $\hat{\phi}_{1,j} = \hat{\phi}_{1}(k, j\kappa)$ and $\hat{\phi}_{2,j} = \hat{\phi}_{2}(k, j \kappa)$, where $\kappa = \Delta t$. Then the solution at $t = (j+1) \Delta t$ is given by
\begin{align}
\hat{\phi}_{1 \lb j+1 \rb \kappa} &= \hat{\phi}_{1, j \kappa} + \frac{1}{6} \lb a_{1} + 2 b_{1} + 2 c_{1} + d_{1} \rb, \notag \\
\hat{\phi}_{2, \lb j+1 \rb \kappa} &= \hat{\phi}_{2, j \kappa} + \frac{1}{6} \lb a_{2} + 2 b_{2} + 2 c_{2} + d_{2} \rb,
\label{PSRK4Coup}
\end{align}
where $a_i$, $b_i$, $c_i$, $d_i$ are functions of $k$ at a given moment in time, $t$, and are defined as
\begin{align*}
&a_{i} = \kappa F_{i} \lb \hat{\phi}_{1, j}, \hat{\phi}_{2, j} \rb, &&b_{i} = \kappa F_{i} \lb \hat{\phi}_{1, j} + \frac{a_{1}}{2}, \hat{\phi}_{2, j} + \frac{a_{2}}{2} \rb, \\
&c_{i} = \kappa F_{i} \lb \hat{\phi}_{1, j} + \frac{b_{1}}{2}, \hat{\phi}_{2, j} + \frac{b_{2}}{2} \rb, &&d_{i} = \kappa F_{i} \lb \hat{\phi}_{1, j} + c_{1}, \hat{\phi}_{2, j} + c_{2} \rb,
\end{align*}
for $i=1, 2$. The functions $F_{i}$ are found as a rearrangement of \eqref{PScOstFFT} to contain all non-time derivatives. Explicitly we have
\begin{align*}
F_{1} \lb \hat{\phi}_{1, j}, \hat{\phi}_{2, j} \rb =& - i k s \alpha_1 \mathscr{F} \lset f_{1} \phi_{1} \rset + \lb i k^3 s^3 \beta_1 - i s k \omega_{1} \rb \hat{\phi}_{1, j} - \frac{i \delta}{sk} \lb \hat{\phi}_{1, j} - \hat{\phi}_{2, j} \rb - \frac{i}{sk} \hat{H}_{1, j}, \notag \\
F_{2} \lb \hat{\phi}_{1, j}, \hat{\phi}_{2, j} \rb =& - i k s \alpha_2 \lset f_{2} \phi_{2} \rset + \lb i k^3 s^3 \beta_2 - i s k \omega \rb \hat{\phi}_{2, j} - \frac{i \gamma}{sk} \lb \hat{\phi}_{2, j} - \hat{\phi}_{1, j} \rb - \frac{i}{sk} \hat{H}_{2, j}.
\end{align*}
To obtain a solution at the next step, the functions $a_{i}, b_{i}, c_{i}, d_{i}$, for $i=1,2$, must be calculated in pairs, that is we calculate $a_{1}$ followed by $a_{2}$, then $b_{1}$ followed by $b_{2}$, and so on.


\begin{thebibliography}{99}
\bibitem{S_book}
A.M. Samsonov, \textit{Strain Solitons in Solids and How to Construct Them}, Chapman \& Hall/CRC, Boca Raton, 2001.
\bibitem{P_book}
A.V. Porubov, \textit{Amplification of Nonlinear Strain Waves in Solids}, World Scientific, Singapore, 2003.
\bibitem{PS}
N. Peake, S.V. Sorokin, A nonlinear model of the dynamics of a large elastic plate with heavy fluid loading, \textit{Proc. R. Soc. A } 462
(2006) 2205Ð2224.
\bibitem{IZKS}
D.A. Indejtsev, M.G. Zhuchkova, D.P. Kouzov, S.V. Sorokin, Low-frequency wave propagation in an elastic plate floating on a two-layered fluid, \textit{Wave Motion} 62 (2016) 98-113.
\bibitem{PTE} T. Peets, K. Tamm, J. Engelbrecht, On the role of nonlinearities in the Boussinesq-type wave equations, \textit{Wave Motion} 71 (2017) 113-119.
\bibitem{ADO} Z. Abiza, M. Destrade, R.W. Ogden, Large acoustoelastic effect, \textit{Wave Motion} 49 (2012) 364-374.
\bibitem{ADKM} I.V. Andrianov, V.D. Danishevsky, J.D. Kaplunov and B. Markert, Wide frequency higher-order dynamic model for transient waves in a lattice, In: I.V. Andrianov et al. ed., ``Problems of Nonlinear Mechanics and Physics of Materials", Springer, 2019.
\bibitem{OS}
L.A. Ostrovsky, A.M. Sutin, Nonlinear elastic waves in rods, \textit{PMM} 41 (1977) 531-537.
\bibitem{NS} G.A. Nariboli, A. Sedov, Burgers-Korteweg de Vries equation for viscoelastic rods and plates, \textit{J. Math. Anal. Appl.} 32 (1970) 661-677.
\bibitem{DF} H.-H. Dai, X. Fan, Asymptotically approximate model equations for weakly nonlinear long waves in compressible elastic rods and their comparisons with other simplified model equations, \textit{Maths. Mechs. Solids} 9 (2004) 61-79.
\bibitem{EKS} V.I. Erofeev, V.V. Kazhaev, N.P. Semerikova, \textit{Waves in rods: dispersion, dissipation, nonlinearity}, Fizmatlit, Moscow, 2002 (in Russian).
\bibitem{GKS} F.E. Garbuzov, K.R. Khusnutdinova, I.V. Semenova, On Boussinesq-type models for long longitudinal waves in elastic rods, arXiv:1810.07684v1 [nlin.PS] 17 Oct 2018 (submitted to \textit{Wave Motion}).
\bibitem{KS}
K.R. Khusnutdinova, A.M. Samsonov, Fission of a longitudinal strain solitary wave in a delaminated bar, \textit{Phys. Rev. E} 77 (2008) 066603.
%\bibitem{DKSS10}
%G.V. Dreiden, K.R. Khusnutdinova, A.M. Samsonov, and I.V. Semenova, Splitting induced generation of soliton trains in layered waveguides, \textit{J. Appl. Phys.}, 107 %(2010) 034909.
\bibitem{DKSS12}
G.V. Dreiden, K.R. Khusnutdinova, A.M. Samsonov, and I.V. Semenova, Bulk strain solitary waves in bonded layered polymeric bars with delamination, \textit{J. Appl. Phys.} 112 (2012) 063516.
\bibitem{KT15}
K.R. Khusnutdinova, M.R. Tranter, Modelling of nonlinear wave scattering in a delaminated elastic bar, \textit{Proc. R. Soc. A} 471 (2015) 20150584.
\bibitem{KT17}
K.R. Khusnutdinova, M.R. Tranter, On radiating solitary waves in bi-layers with delamination and coupled Ostrovsky equations, \textit{Chaos} 27 (2017) 013112.
\bibitem{BBS} A.V. Belashov, Y.M. Beltukov, I.V. Semenova, Pump-probe digital holography for monitoring of long bulk nonlinear strain waves in solid waveguides, \textit{Proc. SPIE} 10678 (2018) 1067810.
\bibitem{KSZ}
K.R. Khusnutdinova, A.M. Samsonov and A.S. Zakharov, Nonlinear layered lattice model and generalized solitary waves in imperfectly bonded structures, \textit{Phys. Rev. E} 79 (2009) 056606.
\bibitem{GKM}
R.H.J. Grimshaw, K.R. Khusnutdinova and K.R. Moore, Radiating solitary waves in coupled Boussinesq equations, \textit{IMA J. Appl. Math.} 82 (2017) 802-820.
\bibitem{KM}
K.R. Khusnutdinova and K.R. Moore, Initial-value problem for coupled Boussinesq equations and a hierarchy of Ostrovsky equations, \textit{Wave Motion} 48 (2011) 738-752.
\bibitem{KT18}
K.R. Khusnutdinova, M.R. Tranter, D'Alembert-type solution of the Cauchy problem for a Boussinesq-type equation with the Ostrovsky term, arXiv:1808.08150v1 [nlin.PS] 24 Aug 2018 (submitted to \textit{Stud. Appl. Math.}).
\bibitem{KMP}
K.R. Khusnutdinova, K.R. Moore and D.E. Pelinovsky, Validity of the weakly nonlinear solution of the Cauchy problem for the Boussinesq-type equation, \textit{Stud. Appl. Math.} 133 (2014) 52-83.
\bibitem{KM12} K.R. Khusnutdinova and K.R. Moore, Weakly non-linear extension of d'Alembert's formula, \textit{IMA J. Appl. Math.} 77 (2012) 361-381.
\bibitem{T}L.N. Trefethen, \textit{Spectral Methods in MATLAB}, SIAM, Philadelphia, 2000.
\bibitem{FFTW3} M. Frigo and S.G. Johnson, The Design and Implementation of FFTW3, \textit{Proc. IEEE} 93 (2005) 216-231.
\bibitem{EST}
J. Engelbrecht, A. Salupere and K. Tamm, Waves in microstructured solids and the Boussinesq paradigm, \textit{Wave Motion} 48 (2011) 717-726.
\bibitem{AGK}
A. Alias, R.H.J. Grimshaw and K.R. Khusnutdinova, On strongly interacting internal waves in a rotating ocean and coupled Ostrovsky equations, \textit{Chaos} 23 (2013) 023121.
\end{thebibliography}
\end{document}